\documentclass[10pt,final,conference]{IEEEtran}
\IEEEoverridecommandlockouts

\usepackage{amsmath}
\usepackage{amssymb}

\usepackage{booktabs}
\usepackage[noadjust]{cite}
\usepackage[hidelinks,breaklinks]{hyperref}

\usepackage{tikz}
\newcommand{\circnum}[2][black]{\tikz[baseline={([yshift=-0.6ex]char.center)}]{%
\node[shape=circle,fill=#1,text=white,inner sep=0pt,minimum size=0.85em,font=\bfseries\tiny] (char) {#2};}}

\usepackage{subcaption}
\usepackage{setspace}

\usepackage{dblfloatfix}
\usepackage{graphicx}
\usepackage{multirow}

\usepackage[ruled,linesnumbered,noend]{algorithm2e}
\DontPrintSemicolon
\SetAlCapFnt{\small}
\SetKwProg{Fn}{\textbf{fn}}{}{end}

\SetCommentSty{mycommentfont}
\makeatletter
  \newcommand{\removelatexerror}{\let\@latex@error\@gobble}
\makeatother

\newcommand{\MemLLM}{\textsc{MemLLM}\xspace}

\begin{document}

\title{MemLLM: End-to-end In-Storage Acceleration of On-device Retrieval-Augmented Generation using Precomputed KV Caches
\thanks{This work was supported by Samsung Research Funding \& Incubation Center for Future Technology under Project Number SRFC-IT2302-05.}}

\author{
  \IEEEauthorblockN{Seungkwan Kang, Hyungseok Ko, Heemin Kim, Myoungsoo Jung}
  \IEEEauthorblockA{\textit{Computer Architecture and Memory Systems Laboratory}, KAIST \\ Daejeon, Republic of Korea}
}

\maketitle

\begin{abstract}
  Retrieval-augmented generation (RAG) on edge devices enables personalized large language model (LLM) inference while preserving privacy.
However, RAG's inherent length asymmetry makes the prefill phase the dominant bottleneck.
While precomputing the KV caches offline alleviates the prefill cost on server systems, the multi-gigabyte caches over the narrow interconnect create a severe I/O wall for on-device systems.
To mitigate this, we propose \MemLLM, an end-to-end in-storage processing (ISP) architecture that executes the complete attention pipeline based on precomputed KV caches within the SSD.
\MemLLM utilizes per-channel accelerators to fully utilize the aggregated NAND channel bandwidth and accelerates all KV cache-related computations on the storage-side to minimize PCIe transfers.
To efficiently handle the sparse updates required by RAG, \MemLLM introduces a fused recomputation-query kernel paired with a hardware-based filter logic to halve the NAND read volume and eliminate write-back penalties.
Furthermore, \MemLLM's global SRAM-based stateful reduction engine enables a completely DRAM-less dataflow, thereby removing the internal memory bottlenecks and bounding the PCIe egress to an $O(1)$ footprint.
A full-stack software stack allows asynchronous, zero-copy transfers from the ISP to the host and provides transparent execution within modern ML frameworks.
Evaluated across eight real-world RAG datasets and six on-device LLM models,  \MemLLM achieves an average end-to-end speedup of 4.06$\times$ over the GPU baseline.

\end{abstract}

\begin{IEEEkeywords}
In-storage processing, retrieval-augmented generation, large language model inference, KV cache, near-data processing, on-device machine learning
\end{IEEEkeywords}

\section{Introduction}\label{sec:introduction}
Large language models (LLMs) are rapidly migrating from cloud servers to edge on-device systems, driven by the demand for privacy-preserving inference over sensitive data, such as personal medical records~\cite{alkhalaf2024applying,wu2024medicalgraphragsafe,amugongo2025retrieval,moller2020covid,jin2019pubmedqa}, financial documents~\cite{choi2025finder,kim2025optimizing,iaroshev2024evaluating,chen2021finqa,zhu2021tat}, and corporate knowledge bases~\cite{hendrycks2021cuad,ravichander2019question,joshi2017triviaqa}, without dependence on centralized servers~\cite{wang2024mememo,qin2024robust,mahapatra2025storage,loo2025mobilerag}.
Retrieval-augmented generation (RAG~\cite{lewis2020retrieval}) is central to this vision.
By grounding LLM responses in an external knowledge base, RAG eliminates the need for costly fine-tuning and enables real-time, personalized answers on-device.

However, RAG workloads exhibit a fundamental length asymmetry; the retrieved contexts typically span thousands of tokens, while the generated answers remain extremely short, often by a factor of 100--1000$\times$.
This asymmetry shifts the inference bottleneck from the decode phase, which dominates standard LLM serving, to the prefill phase, where the LLM must process the massive retrieved context to build its KV cache.
On resource-constrained edge devices, this prefill-dominated latency profile creates a severe performance barrier for practical on-device RAG deployment, especially for the long-context, short-answer regime.

A promising software-level mitigation is to precompute the KV caches for documents in the knowledge base offline and load them from storage at inference time, replacing the compute-intensive prefill with a storage read~\cite{liu2025lmcacheefficientkvcache,yao2025cacheblend,gim2024prompt,10.1145/3773772}.
On server systems equipped with high-bandwidth storage arrays, this trade of computation for I/O yields substantial speedups.
However, on edge devices, the massive data movement across the narrow PCIe interconnect severely throttles performance.
To make matters worse, the limited memory capacity of edge SoCs cannot accommodate these multi-gigabyte precomputed KV caches alongside the model weights, forcing the KV caches to be completely offloaded to persistent storage~\cite{pan2024instinfer,pan2025instattention,jang2026hilos,ren2025characterizing,holmes2024deepspeedfastgenhighthroughputtextgeneration,pmlr-v202-sheng23a}.
Streaming these precomputed caches from the SSD to the GPU over the limited PCIe bus takes longer than the compute time it was meant to replace, creating an \textit{I/O wall} that negates the benefits of precomputation entirely.

Existing in-storage processing (ISP) architectures offer a partial remedy. Prior works have offloaded decode-phase attention to storage devices to eliminate repeated KV cache transfers~\cite{pan2024instinfer,pan2025instattention,jang2026hilos}, or moved model-weight computation into flash to free GPU memory~\cite{lee2025aif}.
Yet, these approaches fail to address the defining challenge of precomputed-KV-based RAG: \textit{selective recomputation}.
Because independently precomputed KV caches break cross-attention between retrieved documents, a small but critical subset of tokens ($\sim$15\%) must be recomputed using the full model weights to restore generation quality~\cite{yao2025cacheblend,yang2025cacheclip}.
This recomputation presents a highly heterogeneous workload consisting of compute-intensive, dense matrix multiplications (GEMMs), element-wise softmax operations and matrix-vector multiplications (GEMVs).
This exceeds the capabilities of not only existing decode-only ISPs, but conventional accelerator paradigms overall.
Rigid 2D systolic arrays suffer catastrophic underutilization (zero-padding bubbles) when faced with 1D decode vectors or irregular sparse masks, while standard Single Instruction Multiple Data (SIMD) architectures lack the inter-PE data forwarding required for efficient GEMM execution.
Unable to absorb the workload internally, current systems are forced into a \textit{disjoint execution} flow: the entire KV cache must be first read from the storage and hauled to the GPU for recomputation, and read yet again for query attention; streaming multi-gigabyte data across the PCIe bus twice.

To mitigate the issue, we present \textit{MemLLM}, an in-storage processing architecture that eliminates this I/O wall by performing the precomputed-KV attention pipeline entirely within the SSD, without ever transferring the full KV cache to the host.
\MemLLM targets on-device RAG with precomputed document KV caches and selective recomputation under long-context, short-answer workloads; we delineate this scope and its boundaries in Section~\ref{sec:discussion}.
\MemLLM's design is governed by three I/O minimization principles.
\enlargethispage{10pt}

\noindent\textit{Reduce NAND I/O.}
\MemLLM fuses selective recomputation and query attention into a single NAND-read pass via a unified kernel that constructs a combined query matrix ($Q_{fused}$) from the recomputation targets and the user query.
A hardware-based stream filter logic compacts the valid tokens from the irregular recomputation mask into a dense stream, allowing the compute units to process only the necessary tokens without wasting cycles on masking and zero-padding.
This fusion halves the NAND read volume and completely eliminates the write-back penalty, as recomputed KV entries are consumed in situ within the same streaming pass.

\noindent\textit{Utilize aggregated NAND bandwidth.}
Rather than funneling all NAND data through a centralized accelerator bottlenecked by the SSD's internal bus, \MemLLM distributes lightweight Elastic-SIMD accelerators to each NAND channel endpoint.
Each Elastic-SIMD dynamically reconfigures its datapath between broadside mode (parallel GEMV for decode) and cascaded mode (systolic-like forwarding for GEMM during prefill), maintaining high PE utilization across the heterogeneous RAG workload.
This per-channel architecture delivers the full aggregate NAND bandwidth directly to computation without any internal bus contention.

\noindent\textit{Eliminate internal DRAM I/O.}
A central coordinator logic merges per-channel partial attention outputs using a stateful online softmax reduction engine, hosted entirely in the controller's global SRAM.
Rather than repeatedly reading and writing partial sums to the controller's DRAM, \MemLLM leverages RAG's length asymmetry to avoid the internal bandwidth bottleneck entirely.
Because $Q_{fused}$ and the final output $O_{final}$ are small enough to reside completely on-chip, the SSD's DRAM is entirely removed from the attention critical path.
To this end, the total PCIe traffic for the entire attention computation is bounded to a single $O(1)$ transfer that is completely independent of context length.

We implemented \MemLLM's Elastic-SIMD architecture in RTL and prototyped it on a 16nm FPGA~\cite{u200}.
We evaluated the cycles required for each GPU kernel using the prototype and an NVIDIA Jetson Orin NX board~\cite{jetsonorinnx} and integrated it with a cycle-accurate SSD simulator~\cite{jung2017simplessd}.
Evaluation across eight real-world datasets and six LLM models shows that \MemLLM achieves an average end-to-end speedup of 4.06$\times$.

\begin{figure}
  \centering
  \vspace{0pt}
  \includegraphics[width=.9\linewidth]{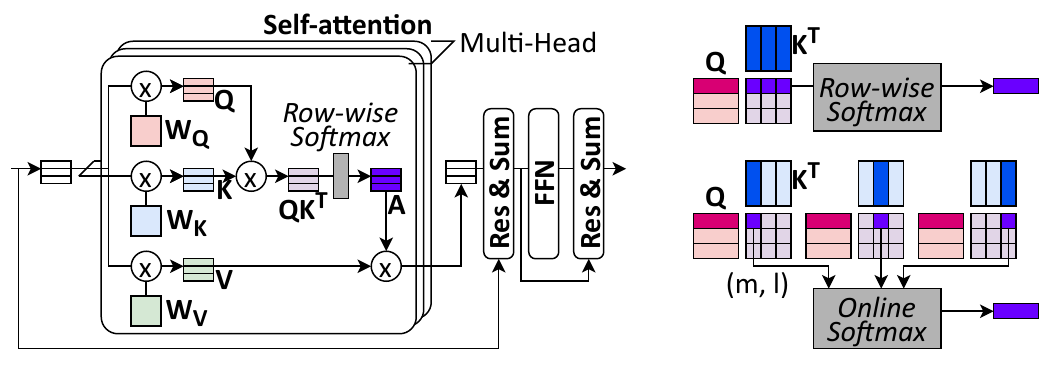}
  \vspace{5pt}
  \caption{LLM inference.}\label{fig:bg_llm}
  \vspace{-32pt}
  \begin{subfigure}{1\linewidth}
    \renewcommand*{\arraystretch}{0.3}
    \begin{tabular}{
      p{\dimexpr.60\linewidth-1.33\tabcolsep}
      p{\dimexpr.35\linewidth-1.33\tabcolsep}
      }
      \caption{Decoder-only transformer.}\label{fig:back_transformer1} &
      \caption{Online softmax.}\label{fig:back_transformer2}
    \end{tabular}
  \end{subfigure}
  \vspace{-5pt}
\end{figure}

The main contributions of this paper are as follows:

\noindent $\bullet$ \textit{Analysis of the on-device RAG I/O wall.}
We provide the first systematic characterization of how precomputed KV caches and their selective recomputation create a disjoint execution bottleneck that conventional accelerator paradigms cannot resolve, establishing the need for an end-to-end ISP for precomputed-KV RAG.

\noindent $\bullet$ \textit{Fused kernel with hybrid mask handling.}
We propose a single-pass fused kernel that eliminates redundant NAND reads and write-backs by co-processing selective recomputation and query attention against the same KV stream, with a hardware-based filter logic that compacts irregular sparse masks into dense compute.

\noindent $\bullet$ \textit{Per-channel Elastic-SIMD.}
We design a distributed, dynamically reconfigurable SIMD fabric that sustains high utilization across GEMV, thin-GEMM, and element-wise softmax workloads without the area overhead and underutilization of rigid 2D systolic arrays.

\noindent $\bullet$ \textit{NAND-stationary tiling and DRAM-less data flow.}
We co-design the hardware micro-architecture with a NAND-stationary tiling algorithm and a global SRAM-based stateful reduction engine that completely bypasses the SSD's DRAM, ensuring that internal storage bandwidth is strictly dedicated to NAND reads rather than intermediate memory traffic.

\section{Background}\label{sec:background}
\subsection{Large Language Model (LLM) Inference}

\noindent\textbf{Decoder-only transformers and KV cache.}
Modern large language models typically employ the decoder-only transformer architecture~\cite{vaswani2017attention}, transforming input token embeddings into context-enriched representations via self-attention and feed-forward network (FFN) layers (Figure~\ref{fig:back_transformer1}).
The core of this architecture is the \textit{scaled dot-product attention}, which projects inputs into query ($Q$), key ($K$), and value ($V$) matrices to compute $A = \text{softmax}(QK^T / \sqrt{d_k})V$.
This captures the contextual relationships by weighting each token's value proportionally to its relevance to the query.
Multi-Head Attention (MHA~\cite{cordonnier2020multi}) further keeps separate $K$/$V$ matrices per $Q$ head and inflating memory bandwidth.
Recent models adopt Grouped-Query Attention (GQA~\cite{ainslie-etal-2023-gqa}), sharing one $K$/$V$ pair per head group to compress the KV cache.

LLM inference operates in two phases.
The \textit{prefill phase} processes the entire input prompt in parallel via dense matrix-matrix multiplications (GEMM).
The subsequent \textit{decode phase} is autoregressive, inferring and appending one token per iteration.
Due to this autoregressive nature, recomputing the full attention for all previously generated tokens incurs a redundant $O(n^2)$ computational cost, where $n$ is the length of the token sequence.
To mitigate this, frameworks utilize the \textit{KV cache}~\cite{vaswani2017attention}.
By storing the $K$ and $V$ vectors of all preceding tokens in memory, the decode phase is reduced to a linear $O(N)$ matrix-vector multiplication (GEMV).
However, this introduces a severe memory bottleneck; for long sequences, the KV cache footprint grows linearly, quickly exhausting the GPU's on-chip VRAM and necessitating storage-level offloading on resource-constrained edge devices.

\begin{figure}
  \centering
  \vspace{0pt}
  \includegraphics[width=1\linewidth]{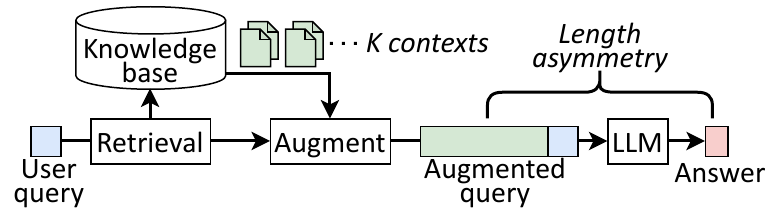}
  \vspace{-15pt}
  \caption{Retrieval-Augmented Generation (RAG).}\label{fig:bg_rag}
  \vspace{-10pt}
\end{figure}

\noindent\textbf{Online Softmax.}
When $QK^T$ is too large to materialize in on-chip SRAM, accelerators compute attention in tiles using the \textit{online softmax} technique~\cite{dao2022flashattention,dao2023flashattention} (Figure~\ref{fig:back_transformer2}), maintaining two per-row running statistics: the maximum value ($m$) and the sum of exponentials ($\ell$).
Partial tile outputs are then merged by rescaling with the exponential difference between local and global maxima, enabling long attentions within a fixed SRAM budget without accuracy loss.

\subsection{Retrieval-Augmented Generation (RAG)}

Retrieval-augmented generation mitigates the hallucination and knowledge-cutoff limitations of LLMs by grounding inference in external, semi-static knowledge bases.
As illustrated in Figure~\ref{fig:bg_rag}, when a user query arrives, a retriever performs a similarity search over a pre-indexed vector database to extract the top-$k$ most relevant passages. These contexts are concatenated with the query to form an augmented prompt for the LLM.
In this work, we assume that the retrieved chunks are finalized before attention begins~\cite{malkov2020efficient,douze2024faiss}. Further evaluation under retrieval methods that relax this assumption~\cite{edge2024local,zhuang2025linearrag,nogueira2020document,trivedi2023interleaving} are in Section~\ref{subsec:eval_sensitivity}.
Unlike standard conversational LLM serving where the decode phase is often the primary bottleneck, RAG workloads exhibit a drastic \textit{length asymmetry}: the retrieved contexts typically span thousands of tokens, while the generated answers remain relatively short.
Consequently, the prefill phase over these massive contexts dominates the end-to-end RAG inference latency (Section~\ref{sec:motivation}).

\subsection{Precomputed KV Caches for RAG}

Because RAG knowledge bases are mostly \textit{static}, the compute-intensive prefill cost for the documents can be amortized.
The system can tokenize each document offline, run a prefill pass to generate its KV cache, and store it in persistent storage~\cite{liu2025lmcacheefficientkvcache,yao2025cacheblend}.
At inference time, instead of recomputing the context from raw tokens, the LLM simply loads the \textit{precomputed KV cache}.
While precomputing KV caches independently per document saves prefill compute, it breaks the \textit{cross-attention} between different retrieved documents. To restore inference accuracy, recent frameworks (e.g., CacheBlend~\cite{yao2025cacheblend}) employ \textit{selective recomputation}.
By evaluating the cross-attention deviations at the first decoder layer, the system flags a small subset ($\sim$15\%) of tokens, termed as High KV Deviation (HKVD) tokens, that require full attention recomputation using the model weights, while the remaining are consumed directly from the storage.

\begin{figure}
  \centering
  \vspace{0pt}
  \includegraphics[width=1\linewidth]{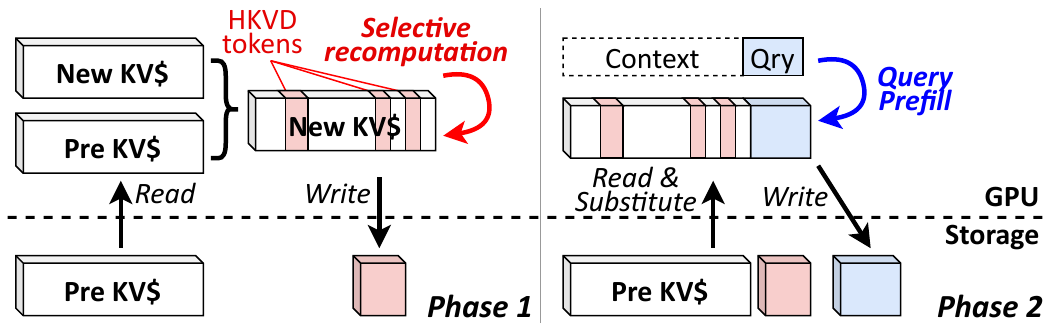}
  \vspace{-12pt}
  \caption{HKVD recomputation.}\label{fig:bg_recomp}
  \vspace{-15pt}
\end{figure}

However, executing this selective recomputation introduces a highly inefficient, disjoint execution flow, as illustrated in Figure~\ref{fig:bg_recomp}.
First, the host GPU must load the entire precomputed KV cache from storage to start the recomputation (Phase 1).
Because only a small portion of the tokens require updates, the $Q$ matrix is a "thin" dense matrix composed strictly of the embeddings for these targeted tokens.
However, because these target tokens are scattered unpredictably across the absolute sequence positions, computing their attention scores requires an irregular, jagged attention mask to maintain positional correctness.
Subsequently, for the query prefill, the host must re-read the entire precomputed KV cache to compute the attention against the newly appended user query (Phase 2).
This structural separation of the KV caches and compute forces the system to stream the multi-gigabyte KV cache across the PCIe bus twice, creating a massive I/O bottleneck that severely limits the efficiency of on-device RAG deployments.

\section{Challenge and Motivation}\label{sec:motivation}
Although on-device RAG promises privacy-preserving, low-latency inference over personalized knowledge bases, its long contexts create severe performance challenges on resource-constrained edge devices.
In this section, we systematically analyze these challenges and derive the architectural requirements for an on-device RAG accelerator.

\subsection{The I/O Wall of Precomputed KV Caches}\label{subsec:motiv_overall}

\noindent\textbf{Datasets and models.}
To ground our analysis in realistic RAG workloads, we collect publicly available question-answering datasets from four domains where RAG is actively deployed: finance (FinQA~\cite{chen2021finqa}, TAT-QA~\cite{zhu2021tat}), biomedical (CovidQA~\cite{moller2020covid}), legal (CuAD~\cite{hendrycks2021cuad}, PrivacyQA~\cite{ravichander2019question}), and general knowledge (TriviaQA~\cite{joshi2017triviaqa}, HotpotQA~\cite{yang2018hotpotqa}).
Table~\ref{tbl:datasets} summarizes their key characteristics.
Total size denotes each dataset's entire precomputed KV cache size under SmolLM2-360M, the medium-sized model in our evaluation.
Most datasets are under 1\,TB and can be readily stored on a commodity SSD.
In contrast, HotpotQA and TriviaQA-wiki, whose corpora are built from the entire Wikipedia, and TriviaQA-web, built from broad web-search documents, are large-scale general-web datasets on the order of several to tens of TB and thus exceed edge storage capacity.
However, these are not intended for edge deployment; we use them to analyze the generality and scalability of our approach across corpus scales.
For these datasets, the evaluation was conducted based on access-pattern traces rather than directly storing the full KV cache.

\begin{figure}
  \centering
  \vspace{0pt}
  \includegraphics[width=1\linewidth]{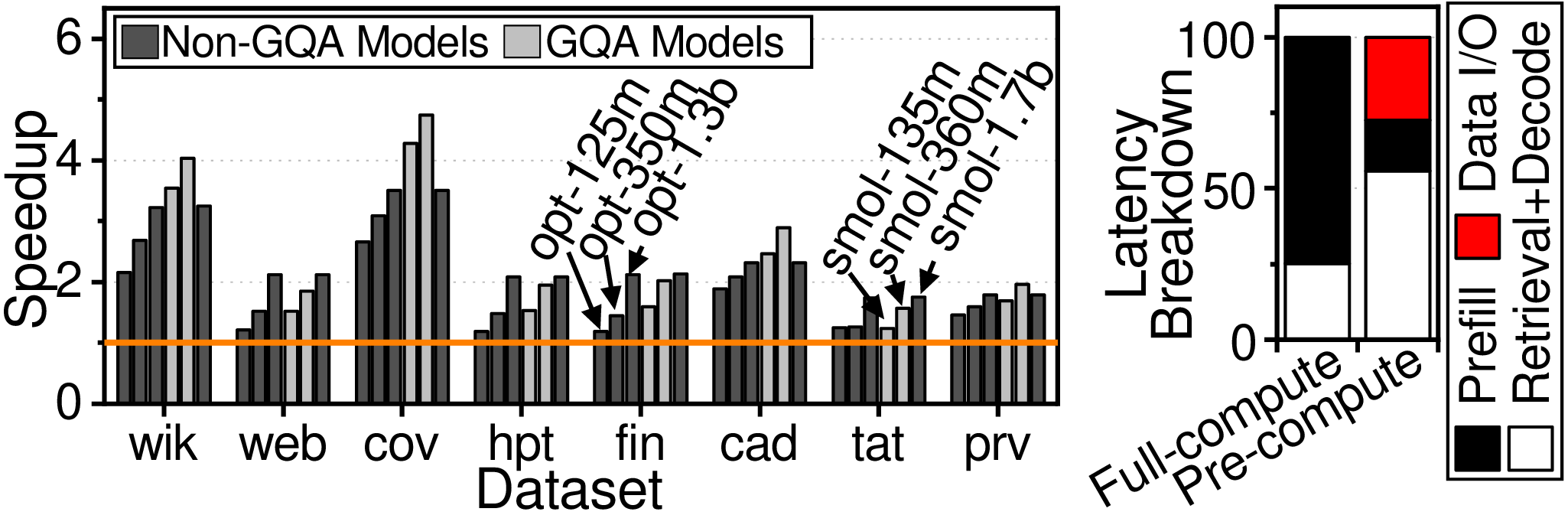}
  \vspace{-5pt}
  \caption{Precomputed KV caches on server systems.}\label{fig:motiv-server-merged-colored}
  \vspace{-30pt}
  \begin{subfigure}{1\linewidth}
    \renewcommand*{\arraystretch}{0.3}
    \begin{tabular}{
      p{\dimexpr.68\linewidth-1.33\tabcolsep}
      p{\dimexpr.32\linewidth-1.33\tabcolsep}
      }
      \caption{Normalized speedup.}\label{fig:motiv-server-merged-colored1} &
      \caption{Breakdown.}\label{fig:motiv-server-merged-colored2}
    \end{tabular}
  \end{subfigure}
\end{figure}

\begin{table}[b]
  \small
  \renewcommand*{\arraystretch}{0.3}
  \setlength{\tabcolsep}{2.6pt}
  \resizebox{\linewidth}{!}{%
  \begin{tabular}{ccccccc}
  \toprule
  \multirow{2}{*}{\textbf{Label}} & \multirow{2}{*}{\textbf{Dataset}} & \textbf{\# of}    & \textbf{Context} & \textbf{Answer}   & \multirow{2}{*}{\textbf{Ctx/Ans}} & \textbf{Total}     \\
                                  &                                   & \textbf{Contexts} & \textbf{Length}  & \textbf{(tokens)} &                                   & \textbf{Size} \\
  \midrule
  tat & TAT-QA        & 2.76k & 364   & 5.49 & 66.4 & 38GB   \\
  fin & FinQA         & 2.79k & 688   & 4.18 & 165  & 73GB   \\
  hpt & HotpotQA      & 113k  & 1.11k & 2.35 & 472  & 4.7TB  \\
  web & TriviaQA-web  & 529k  & 2.10k & 4.37 & 479  & 41.4TB \\
  wik & TriviaQA-wiki & 111k  & 4.33k & 4.34 & 997  & 17.9TB \\
  prv & PrivacyQA     & 1.75k & 4.34k & 187  & 23.2 & 290GB  \\
  cov & CovidQA       & 642   & 6.12k & 13.9 & 440  & 150GB  \\
  cad & CuAD          & 510   & 13.7k & 151  & 90.7 & 267GB  \\
  \bottomrule
  \end{tabular}
  }
  \caption{RAG datasets.}\label{tbl:datasets}
  \vspace{-10pt}
\end{table}

Across all domains, we observe a pronounced length asymmetry: the contexts typically span thousands of tokens, while the answers range from a few tokens to at most a few hundred tokens.
This asymmetry reflects the primary use case of RAG: extracting specific factual information from long documents.
For example, TriviaQA retrieves long documents on general knowledge as context, but produces short factual answers, resulting in a context-to-answer ratio of approximately 1000:1.
Legal contract analysis (CuAD) and biomedical QA (CovidQA) also exhibit ratios of approximately 90:1 and 440:1, respectively.
This asymmetry determines the latency profile of RAG inference: the long contexts inflate prefill time, while the short answers limit the number of decode iterations.

For LLM models, we evaluated OPT~\cite{zhang2022opt} (350M to 6.7B) and SmolLM2~\cite{allal2025smollm2smolgoesbig} (135M to 1.7B), spanning a range of parameter counts representative of on-device deployment.
These model families include both grouped-query attention (GQA) and non-GQA variants, enabling us to analyze the impact of KV cache sizes on I/O overhead.
The key characteristics of the models are summarized in Table~\ref{tbl:models}.
All models were evaluated in FP16 precision on an NVIDIA Jetson Orin~\cite{jetsonorinnx} system, representative of on-device edge environments.
Hardware and software details are summarized in Table~\ref{tbl:setup}.

\begin{figure}
  \centering
  \vspace{0pt}
  \includegraphics[width=1\linewidth]{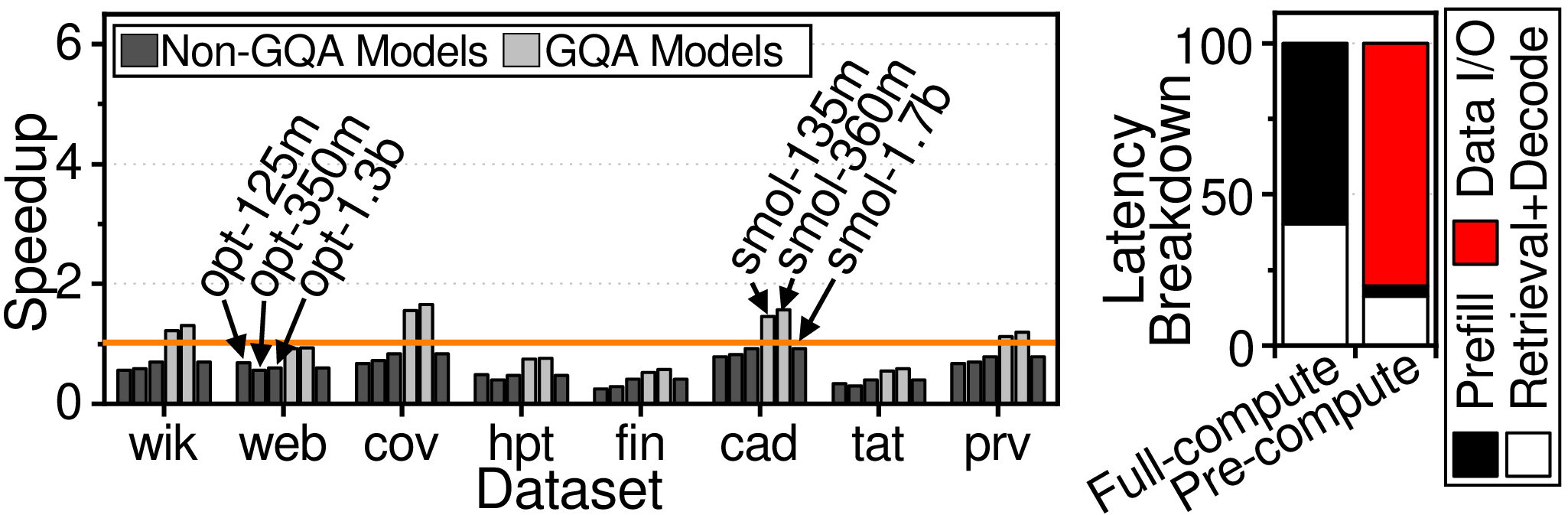}
  \vspace{-5pt}
  \caption{Precomputed KV caches on on-device systems.}\label{fig:motiv-on-device-merged-colored}
  \vspace{-30pt}
  \begin{subfigure}{1\linewidth}
    \renewcommand*{\arraystretch}{0.3}
    \begin{tabular}{
      p{\dimexpr.68\linewidth-1.33\tabcolsep}
      p{\dimexpr.32\linewidth-1.33\tabcolsep}
      }
      \caption{Normalized speedup.}\label{fig:motiv-on-device-merged-colored1} &
      \caption{Breakdown.}\label{fig:motiv-on-device-merged-colored2}
    \end{tabular}
  \end{subfigure}
\end{figure}

\noindent\textbf{Server and on-device system comparison.}
Precomputing the documents' KV caches replaces the compute-intensive context prefill with a storage read, trading computation for I/O.
On servers with high-bandwidth storage and interconnects, this yields an average end-to-end speedup of 2.1$\times$ over the full-compute baseline across eight RAG datasets (Figure~\ref{fig:motiv-server-merged-colored1}): the prefill fraction drops from 74.8\% to 16.9\%, with the storage read accounting for only 27.4\% of total latency (Figure~\ref{fig:motiv-server-merged-colored2}).
On the Jetson Orin (Figure~\ref{fig:motiv-on-device-merged-colored}), the average end-to-end speedup instead drops below 0.58$\times$: unlike the server interconnect, which delivers precomputed KV caches faster than the GPU can compute them, the edge device's limited PCIe bandwidth cannot, so an \textit{I/O wall} negates the savings of precomputation.
Figure~\ref{fig:motiv-on-device-merged-colored2} confirms this: the prefill compute fraction falls from 59.8\% to 3.53\% as expected, but the storage I/O fraction balloons to 80.2\% of total latency, the new dominant bottleneck.
Models with grouped-query attention (GQA) produce smaller KV caches and thus see only marginal on-device speedups, yet I/O still consumes a substantial fraction of their prefill latency, showing that model-level reduction alone cannot fundamentally overcome the I/O wall.

\begin{table}[b]
  \small
  \renewcommand*{\arraystretch}{0.3}
  \setlength{\tabcolsep}{3.8pt}
  \resizebox{\linewidth}{!}{%
  \begin{tabular}{ccccccc}
  \toprule
  \textbf{Model}  & \textbf{\# of}   & \multirow{2}{*}{\textbf{GQA}} & \multirow{2}{*}{\textbf{Layers}} & \textbf{Hidden} & \textbf{KV\$/token} & \textbf{Max tokens} \\
  \textbf{Family} & \textbf{Params.} &                               &                               & \textbf{Dimension} &\textbf{(KB)}    & \textbf{in 8GiB}    \\
  \midrule
  \multirow{3}{*}{OPT}    & 125M &   & 12 & 768  & 32 & 127k \\
                          & 350M & X & 24 & 1024 & 96 & 40.0k \\
                          & 1.3B &   & 24 & 2048 & 192 & 14.9k \\
  \midrule
  \multirow{3}{*}{SmolLM} & 135M & 1:3 & 30 & 576  & 22.5 & 180k \\
                          & 360M & 1:3 & 32 & 960  & 40   & 95.6k \\
                          & 1.7B &  X  & 24 & 2048 & 192  & 12.8k \\
  \bottomrule
  \end{tabular}
  }
  \caption{LLM models.}\label{tbl:models}
  \vspace{-10pt}
\end{table}

\noindent\textbf{KV cache offloading.}
The analysis so far assumed the KV caches fit in GPU memory (we shrank the layer count to fit on-device memory and extrapolated the measured latency to full-model scale), but this rarely holds in practice.
A model's footprint is its own weights plus the KV cache, whose size grows with the per-token KV size, the number of retrieved contexts (top-$k$), and the dataset's average context length.
Table~\ref{tbl:models} reports the maximum tokens each model fits in 8GB: even \texttt{OPT-350M} tops out at 40k tokens, too few for top-4 retrieval on CuAD, and at a moderate $k{=}5$ most model-dataset pairs exceed the budget.
Consequently, for a significant fraction of practical on-device RAG workloads, the precomputed KV caches must reside in storage rather than GPU memory.
This storage-resident regime, which we call \textit{KV cache offloading}, is critical for on-device RAG as it amplifies the I/O wall, and is the focus of the rest of this section.

\subsection{Analysis of End-to-End On-Device RAG}\label{subsec:motiv_e2e}

\begin{figure}
  \centering
  \vspace{0pt}
  \begin{minipage}[t]{0.274\linewidth}
    \vspace{0pt}
    \includegraphics[width=1\linewidth]{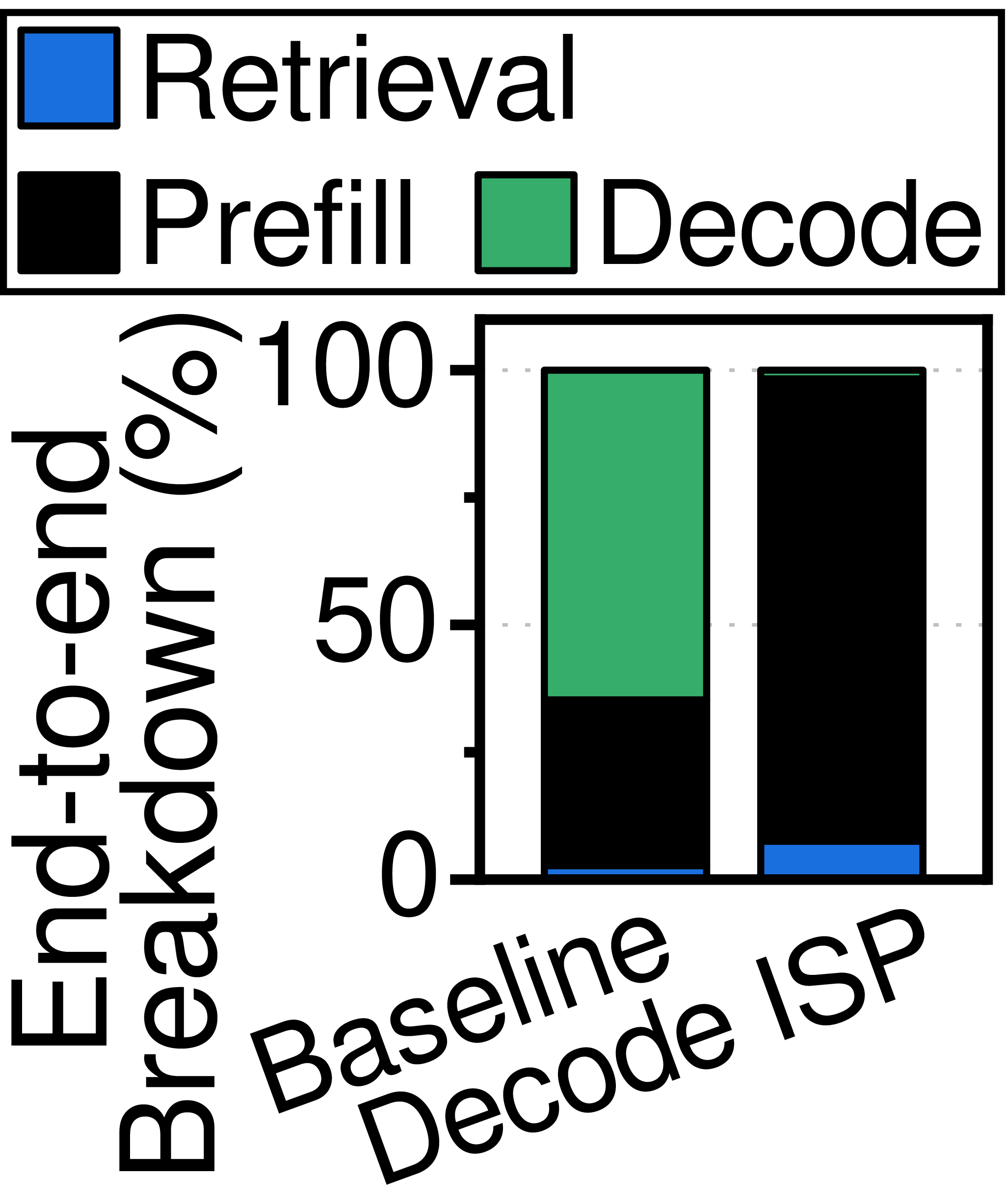}
    \vspace{-12pt}
    \caption{End-to-end breakdown.}\label{fig:motiv-e2e-breakdown}
  \end{minipage}%
  \begin{minipage}[t]{0.019\linewidth}
    \vspace{0pt}
    \hspace{\linewidth}
  \end{minipage}%
  \begin{minipage}[t]{0.343\linewidth}
    \vspace{0pt}
    \includegraphics[width=1\linewidth]{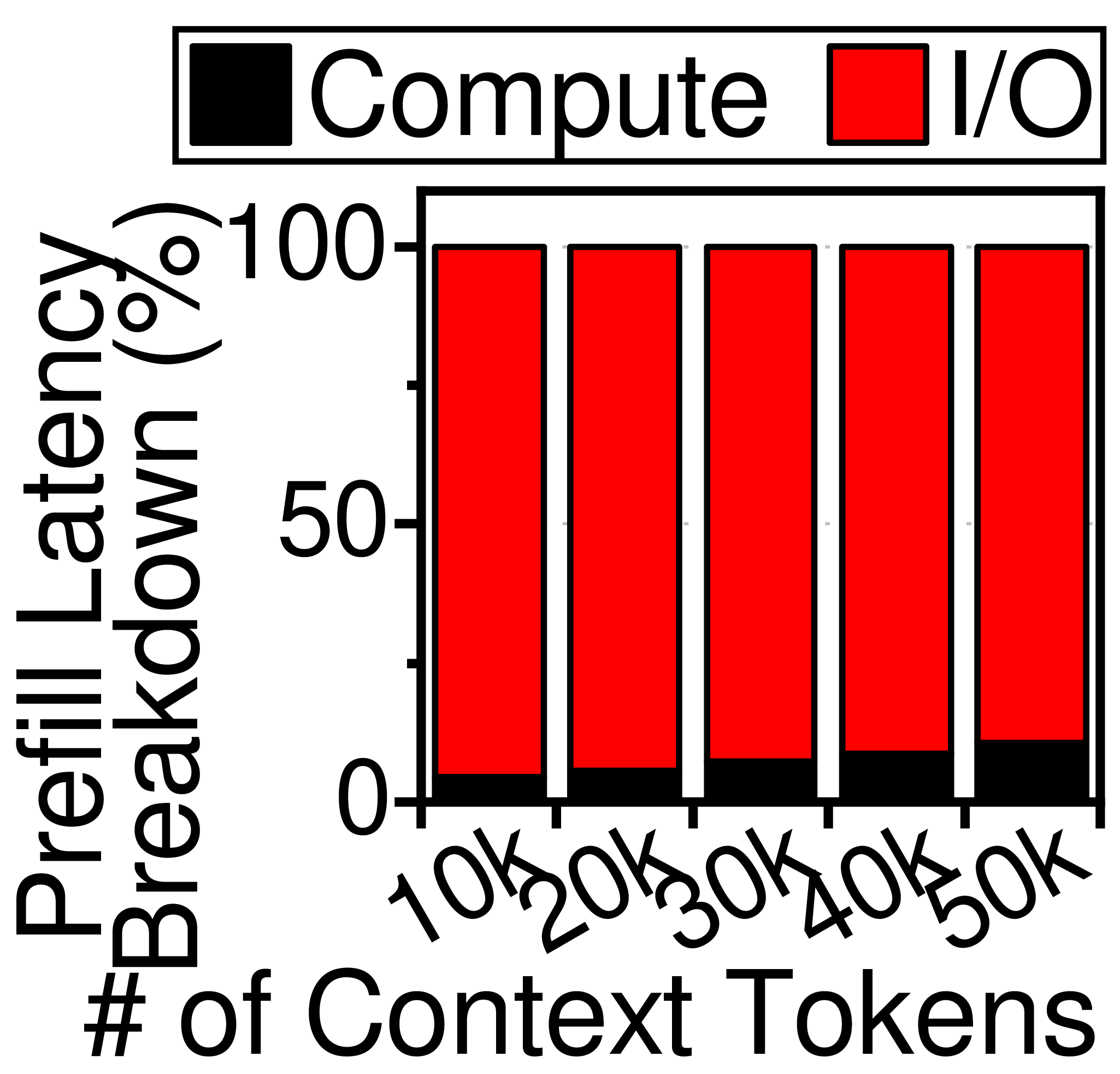}
    \vspace{-12pt}
    \caption{Prefill latency breakdown.}\label{fig:motiv-prefill-breakdown}
  \end{minipage}%
  \begin{minipage}[t]{0.016\linewidth}
    \vspace{0pt}
    \hspace{\linewidth}
  \end{minipage}%
  \begin{minipage}[t]{0.353\linewidth}
    \vspace{0pt}
    \hspace{\linewidth}
    \includegraphics[width=1\linewidth]{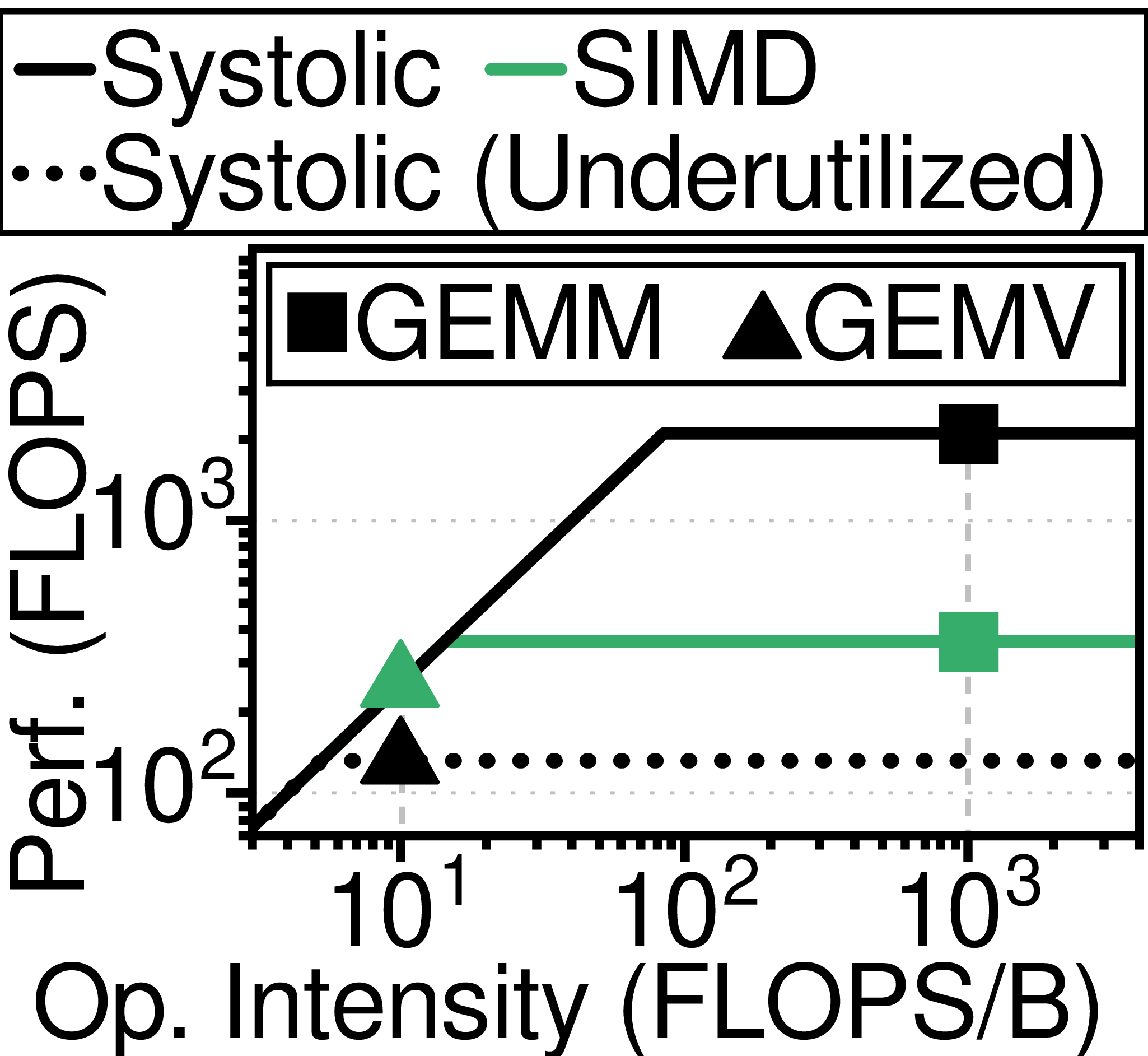}
    \vspace{-10pt}
    \caption{Roofline.}\label{fig:motive-roofline}
  \end{minipage}%
  \vspace{-10pt}
\end{figure}

\noindent\textbf{Decomposing the end-to-end latency.}
When KV caches are offloaded to storage, both phases of RAG inference become I/O-bound.
Decode streams the entire multi-gigabyte KV cache over the bandwidth-limited PCIe at every token step, dominating the baseline at over 64\% of total latency (Figure~\ref{fig:motiv-e2e-breakdown}, left).
Prior in-storage processing (ISP) work offloads decode-phase attention to storage~\cite{pan2025instattention,jang2026hilos,kim2025llmonthepalm}, but even an ideal decode-only ISP (accessing KV at full aggregate NAND bandwidth and matching the GPU's FLOPS) merely shifts the bottleneck: prefill latency then inflates and negates the E2E speedup (Figure~\ref{fig:motiv-e2e-breakdown}, right).
The root cause is RAG's selective recomputation of the precomputed KV caches.
Because it requires GEMMs beyond a decode-only ISP, recomputation must run on the host, forcing a PCIe round-trip during prefill: the precomputed KV is hauled to the GPU, recomputed, written back, and re-read for the query prefill.
This disjoint execution is fundamentally I/O-bound, with I/O exceeding 96\% of prefill time regardless of context length (Figure~\ref{fig:motiv-prefill-breakdown}).

Accelerating on-device RAG therefore requires an \textit{end-to-end ISP} that absorbs both prefill (recomputation and query attention) and decode entirely within storage, and, within the $\sim$3W storage power budget, a flexible microarchitecture that handles RAG's heterogeneous execution profile without intermediate PCIe egress.

\noindent\textbf{Roofline analysis.}
The key challenge for such an end-to-end ISP is that the RAG inference pipeline presents a heterogeneous mix of operators with widely varying arithmetic intensities, all of which must execute within the power budget.
Figure~\ref{fig:motive-roofline} presents a roofline analysis of these major operators mapped onto two conventional accelerator architectures: 2D systolic arrays and 1D SIMD arrays.
For standard, dense GEMMs, systolic arrays reach near-peak compute utilization.
However, their fixed two-dimensional dataflow is fundamentally mismatched with the rest of the RAG pipeline.
The most fatal degradation occurs during the decode-phase GEMV: when the sequence dimension drops to 1, the vast majority of processing elements (PEs) in a rigid 2D grid sit idle.
Furthermore, processing the thin-GEMMs (due to the smaller sequence dimension of the 15\% recomputation targets) required for selective recomputation exacerbates this underutilization, while handling the irregular attention masks introduces additional boundary zero-padding overhead.
Under a strict ~3W budget, dedicating precious silicon area to a large 2D array that remains heavily underutilized for the majority of the E2E latency is architecturally prohibitive.
Conversely, SIMD-only architectures handle 1D GEMV and element-wise vector operations (e.g., softmax) efficiently, as each PE operates independently, but lack the low-cost inter-PE data forwarding necessary for dense GEMMs.
Consequently, partial sums must be heavily routed through shared memory, creating an internal memory bandwidth bottleneck that severely limits the compute ceiling for projection operators.

Since neither architecture alone covers the full spectrum of RAG operators, this analysis yields two architectural requirements for an on-device RAG ISP.
i) \textit{Elasticity}: The ISP must switch seamlessly between parallel vector operations (GEMV) and matrix operations (GEMM).
ii) \textit{Sparsity-awareness}: The ISP must gracefully filter and compact irregular attention masks without incurring control or padding overheads.
In the following section, we present \MemLLM, an architecture purpose-built to satisfy these requirements.

\section{Overview}\label{sec:overview}
\begin{figure*}
  \centering
  \vspace{0pt}
  \begin{minipage}[t]{0.85\linewidth}
    \vspace{0pt}
    \includegraphics[width=1\linewidth]{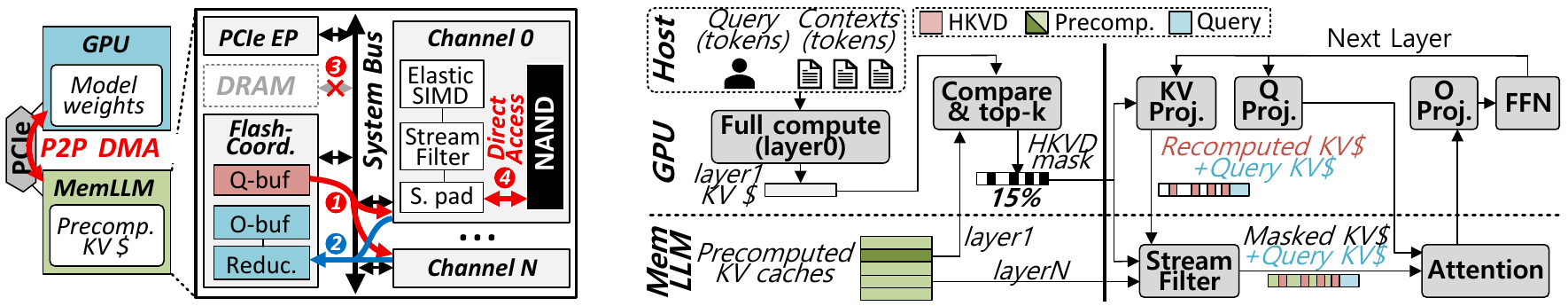}
  \end{minipage}%
  \begin{minipage}[t]{0.02\linewidth}
    \vspace{0pt}
    \hfill
  \end{minipage}%
  \begin{minipage}[t]{0.13\linewidth}
    \vspace{0pt}
    \includegraphics[width=0.96\linewidth]{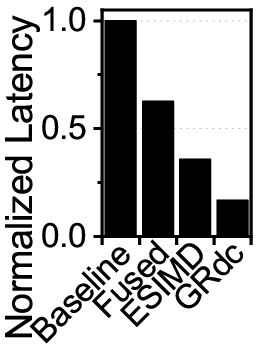}
  \end{minipage}
  \begin{minipage}[t]{0.32\linewidth}
    \vspace{5pt}
    \caption{MemLLM overview.}\label{fig:overview_topology}
    \vspace{-10pt}
  \end{minipage}%
  \begin{minipage}[t]{0.02\linewidth}
    \vspace{0pt}
    \hfill
  \end{minipage}%
  \begin{minipage}[t]{0.45\linewidth}
    \vspace{5pt}
    \caption{Dataflow of fused attention kernel.}\label{fig:overview_dataflow}
    \vspace{-10pt}
  \end{minipage}%
  \begin{minipage}[t]{0.02\linewidth}
    \vspace{0pt}
    \hfill
  \end{minipage}%
  \begin{minipage}[t]{0.22\linewidth}
    \vspace{5pt}
    \caption{Waterfall.}\label{fig:waterfall}
    \vspace{-10pt}
  \end{minipage}%
\end{figure*}

To mitigate the I/O wall of KV caches, we designed \MemLLM, an in-storage processing architecture that accelerates on-device RAG by performing the full attention computation within the SSD, completely eliminating massive KV cache transfers.
Our design is governed by a unified I/O minimization principle applied across three levels of the memory hierarchy: fusing operations to halve NAND read volume, distributing compute to fully exploit aggregate channel bandwidth, and confining partial reductions to on-chip SRAM to bypass the controller's DRAM. This section presents the overall system (Section~\ref{subsec:overview-topology}), traces the end-to-end dataflow of a RAG query (Section~\ref{subsec:overview-e2e}), and quantifies the cumulative performance contribution of each design (Section~\ref{subsec:overview-contribution}).

\subsection{System Overview}\label{subsec:overview-topology}

Figure~\ref{fig:overview_topology} illustrates the \MemLLM system, consisting of a host GPU and a \MemLLM CSSD connected via PCIe. The host GPU retains model weights in its memory and executes projection and FFN layers.
\MemLLM persists all KV caches in flash, including the precomputed caches and the newly computed ones, and performs attention over them.
Communication with the \MemLLM SSD bypasses the conventional block I/O stack.
We adopt an NVMe vendor-specific command to carry tensor metadata and DMA addresses directly to the SSD, establishing a P2P DMA path for zero-copy transfers between the GPU and the ISP (Section~\ref{subsec:software-stack}).

At the center of the \MemLLM SSD, the controller houses the \textit{Flash-Coordinator}, the central scheduling engine that orchestrates multi-channel synchronization and NAND timing. The controller also contains 1MB of global SRAM partitioned into two buffers.
The \textit{Q broadcast buffer} holds the query matrix received from the host and serves it repeatedly to all channels at on-chip bandwidth (\circnum[red]{1}).
The \textit{O reduction buffer} accumulates partial attention outputs from all channels via a stateful reduction engine that performs online softmax correction in place (\circnum[blue]{2}).
The controller's DRAM is entirely excluded from the attention data path; it remains dedicated to Flash Translation Layer (FTL) management, avoiding bandwidth contention and power overheads (\circnum[red]{3}).
Each of the $N$ flash channels features a near-flash accelerator comprising three core components. First, the \textit{Stream Filter} filters invalid tokens against the recomputation mask before data enters the compute pipeline. Second, the \textit{Elastic-SIMD} cluster dynamically reconfigures between GEMV and GEMM execution modes using elastic datapath routing. Finally, a 72KB local scratchpad enables double-buffering of KV pages for latency-hiding via die-level pipelining.
Data flows from NAND pages directly into this per-channel accelerator without traversing any shared internal SSD bus, ensuring that the aggregate channel bandwidth is fully dedicated to computation (\circnum[red]{4}).

\subsection{End-to-End Dataflow}\label{subsec:overview-e2e}

The selective-recomputation-based prefill comprises three steps: i) identifying the HKVD tokens, ii) recomputing the KV cache of these HKVD tokens, and iii) performing prefill for the user query using both the precomputed and recomputed KV caches.
To execute these steps simultaneously without re-reading the KV caches, \MemLLM introduces a \textit{fused attention kernel}.
Figure~\ref{fig:overview_dataflow} illustrates the end-to-end flow of this fused kernel.
At the start of inference, the host uses the query and the retrieved-context tokens to fully compute the first decoder layer (layer 0), thereby obtaining the KV vectors of layer 1.
It then reads the precomputed KV cache of the contexts from storage. Because only the layer-1 KV cache is read at this stage, less than 5\% of the total KV cache is transferred.
The host compares the newly computed KV cache against the precomputed KV cache, computes the token-wise deviation and selects roughly 15\% of the tokens with the largest deviation as HKVD tokens.
The mask of these HKVD tokens is transmitted to \MemLLM to initialize the Stream Filter.
Here, the recomputation ratio (5--18\%~\cite{yao2025cacheblend}) is a static, user-determined hyperparameter that controls the trade-off between generation quality and compute cost.
In the same pass that handles the HKVD recomputation targets, \MemLLM's fused kernel also performs the attention for the new user query; through the iterative computation of subsequent layers, the KV caches of both the HKVD tokens and the query are obtained in a single pass.
To this end, the host recomputes the KV cache of the HKVD tokens selected in the previous layer and transmits it, together with the query's KV cache, to \MemLLM.

\MemLLM then combines this host-supplied data with the precomputed KV cache read from flash utilizing the Stream Filter logic into masked KV cache.
It then performs the attention computation and returns the result to the host; the internal datapath that carries out this computation, comprising channel-parallel tiling, per-channel compute, and DRAM-less on-chip reduction, is detailed in Section~\ref{sec:design}.
At this point, the KV caches received from the host are stored out-of-place in flash and later used for the decode computation. Note that the original precomputed KV cache is preserved unmodified, since it is needed when processing other queries in the future.
Based on the attention result computed by \MemLLM, the host performs the projection and FFN computations to produce the KV cache of the next layer; this process repeats for as many layers as the model has, computing the KV caches of the HKVD tokens and the query tokens across all layers in a single iteration.
Crucially, the bulk precomputed KV cache never leaves the SSD; only the compact $Q_{fused}$, the host-recomputed KV subset, and the returned $O_{final}$ cross the PCIe bus.
The egress thus stays at an $O(1)$ footprint regardless of context length, whereas storage-offloading baselines must repeatedly stream the full $O(N)$ KV cache.

\subsection{Performance Improvement Analysis}\label{subsec:overview-contribution}

Figure~\ref{fig:waterfall} quantifies the cumulative contribution of each microarchitectural design through a waterfall analysis of the end-to-end RAG latency.
The \texttt{Baseline} bar represents the on-device RAG system with precomputed KV caches under storage offloading, as characterized in Section~\ref{subsec:motiv_e2e}.
Starting from the baseline, we apply each design incrementally.
Note that the host-side HKVD identification that constructs the jagged attention mask is a fixed cost embedded in the end-to-end latency of every bar, including the \texttt{Baseline}.
By fusing recomputation and query attention into a single NAND-read pass (\texttt{Fused Kernel}), \MemLLM eliminates the redundant KV re-read, halving the prefill read volume for a 38\% latency reduction.
Distributing Elastic-SIMD accelerators to each channel endpoint further eliminates the internal bus bottleneck, delivering the full aggregate NAND bandwidth directly to the compute units (\texttt{Elastic-SIMD}).
Concurrently, the Stream Filter suppresses the 15\% of invalid tokens before they consume compute cycles.
Coupled with the elastic datapath that maintains high PE utilization across both GEMV and GEMM workloads, this distributed architecture provides an additional 39\% latency reduction.
The Flash-Coordinator's DRAM-less reduction engine merges per-channel partial outputs within global SRAM (\texttt{Global Rdc}), operating at on-chip bandwidth without FTL contention. PCIe traffic is thus bounded to $Q_{fused}$ and $O_{final}$ ($\sim$512KB) regardless of context length, contributing a further 63\% latency reduction and reaching 86\% of the aggregate-NAND-read minimum.
Collectively, these three design principles reduce the end-to-end RAG latency by 7.1$\times$ compared to the baseline.

\section{Design and Implementation}\label{sec:design}
Building upon the I/O minimization principles established in Section~\ref{sec:overview}, this section details the microarchitectural implementation of \MemLLM.
We first present the fused kernel and its internal dataflow, followed by the Elastic-SIMD microarchitecture, the NAND-aware tiling algorithm, the DRAM-less dataflow, and finally the full-stack software integration.

\subsection{Fused Kernel and Internal Dataflow}\label{subsec:fused-kernel}

\begin{figure}
  \centering
  \vspace{0pt}
  \includegraphics[width=1\linewidth]{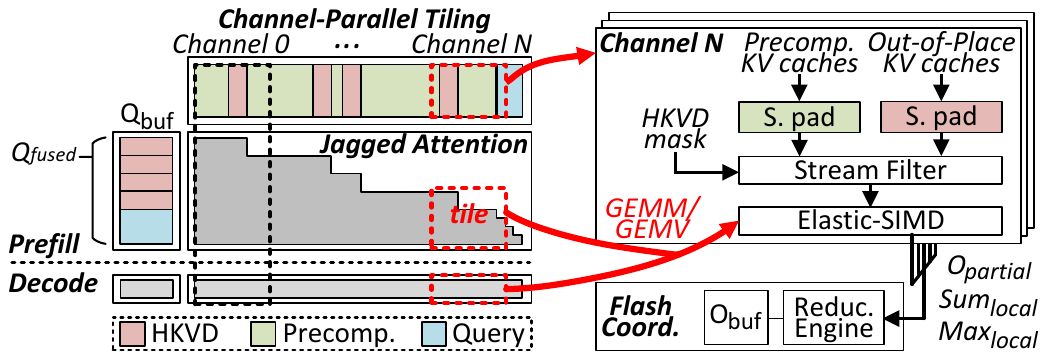}
  \vspace{-2pt}
  \caption{MemLLM internal dataflow.}\label{fig:fused-kernel-mask}
  \vspace{-32pt}
  \begin{subfigure}{1\linewidth}
    \renewcommand*{\arraystretch}{0.3}
    \begin{tabular}{
      p{\dimexpr.5\linewidth-1.33\tabcolsep}
      p{\dimexpr.5\linewidth-1.33\tabcolsep}
      }
      \caption{Channel-parallel tiling.}\label{fig:tiling-left} &
      \caption{Per-channel compute.}\label{fig:tiling-right}
    \end{tabular}
  \end{subfigure}
  \vspace{-0pt}
\end{figure}

\MemLLM executes both the GEMM-based fused attention kernel of prefill and the GEMV-based decode kernel on its per-channel Elastic-SIMD, keeping the bulk KV cache resident in flash and off the PCIe bus throughout inference.
Figure~\ref{fig:fused-kernel-mask} details the internal dataflow of each kernel.

\noindent\textbf{Fused kernel.}
For the fused kernel, the host supplies the query vectors of the HKVD recomputation targets and the new query tokens, which are stacked into a single dense query matrix $Q_{fused}$ and transmitted together with a bitvector we call the \textit{jagged attention mask}.
The kernel scores $Q_{fused}$ against the concatenated $K$ vectors, i.e., the precomputed context $K$, the recomputed HKVD $K$, and the new query $K$, gated by this mask.
The jagged attention mask captures the hybrid causal structure of $Q_{fused}$: the upper rows correspond to a discontinuous subset of original sequence positions (the HKVD targets), so their causal boundaries do not form a standard lower-triangular matrix but shift abruptly with each token's absolute position, producing a irregular, jagged staircase; the lower rows, representing the new query, take a standard dense mask over the entire context.
The Elastic-SIMD applies this mask during the in-situ softmax to guarantee causal correctness.

\noindent\textbf{Channel-parallel tiling.}
Since the KV cache resides per channel in flash, \MemLLM parallelizes this score computation along the $K$ dimension: as Figure~\ref{fig:tiling-left} shows, each channel evaluates the jagged-masked scores for the KV slice it stores, an interleaved mix of precomputed-context and HKVD contexts.
Because every channel must see the full query, the Flash-Coordinator deposits $Q_{fused}$ into the $Q$ broadcast buffer ($Q_{buf}$) and broadcasts it to all channels; a tile then pairs one KV page with a $Q_{fused}$ tile.
The decode kernel reuses the identical tiling with $Q_{fused}$ collapsed to a single query row (GEMV).
Overall, \MemLLM is a faithful hardware realization of CacheBlend\cite{yao2025cacheblend}, not an approximation: its tile-by-tile fused kernel and online-softmax reduction merely reorder otherwise-identical computations, so the output matches a software CacheBlend execution up to floating-point rounding, with negligible downstream accuracy impact (Section~\ref{subsec:eval_overall}).

\begin{figure}
  \centering
  \vspace{0pt}
  \begin{minipage}[t]{1\linewidth}
    \vspace{0pt}
    \includegraphics[width=1\linewidth]{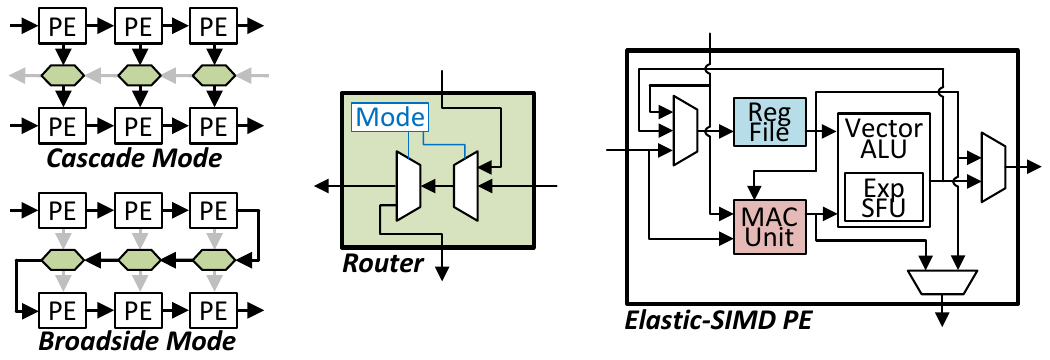}
    \vspace{-3pt}
    \caption{Elastic SIMD.}\label{fig:elastic_simd}
    \vspace{-32pt}
    \begin{subfigure}{1\linewidth}
      \renewcommand*{\arraystretch}{0.3}
      \begin{tabular}{
        p{\dimexpr.26\linewidth-1.33\tabcolsep}
        p{\dimexpr.28\linewidth-1.33\tabcolsep}
        p{\dimexpr.46\linewidth-1.33\tabcolsep}
        }
        \caption{Dual mode.}\label{fig:elastic_simd1} &
        \caption{Router.}\label{fig:elastic_simd2} &
        \caption{Internal PE.}\label{fig:elastic_simd3}
      \end{tabular}
    \end{subfigure}
  \end{minipage}%
\end{figure}

\noindent\textbf{Per-channel compute.}
Within a channel (Figure~\ref{fig:tiling-right}), two scratchpad regions feed the datapath: one buffers the original precomputed KV streamed directly from flash, while the other holds the out-of-place KV, namely the newly written entries: the recomputed HKVD KV, the new query's KV, and the KV of tokens produced during decode.
Guided by the 1D \textit{HKVD mask}, the Stream Filter merges the two into a single dense masked-KV tile, substituting the recomputed entries at the HKVD positions and dropping the masked-out ones, and forwards it to the Elastic-SIMD.
Each channel emits a partial output $O_{partial}$ with its local softmax statistics ($\text{Max}_{local}$, $\text{Sum}_{local}$), which the Flash-Coordinator's reduction engine accumulates in the $O$ reduction buffer ($O_{buf}$) via log-sum-exp correction before the final $O_{final}$ is returned to the host.

\noindent\textbf{Write latency hiding.}
The recomputed HKVD, query, and decode KV that the host ships back are persisted to flash, yet stay off the critical path.%
Appended out of place rather than updated in place, they leave the original precomputed KV untouched and reusable across requests with a different HKVD subset, and incur no read-modify-write.
During the fused pass the attention runs in situ in the Elastic-SIMD registers, freeing the NAND write path, so the controller streams the appends in the background, overlapped with the host GPU's next-layer projection and FFN, and the flash's modest write bandwidth absorbs them (Section~\ref{subsec:eval_overall}). Decode appends only one token's KV per step, which is negligible.

\subsection{Elastic-SIMD Microarchitecture}\label{subsec:elastic_simd}

To execute the heterogeneous RAG pipeline within a strict $\sim$3W storage power envelope, we designed \textit{Elastic-SIMD}, a highly reconfigurable compute fabric situated at each NAND channel endpoint.
Figure~\ref{fig:elastic_simd} details its internal organization.
Unlike rigid 2D systolic arrays, the Elastic-SIMD is a 1D array of Processing Elements (PEs) -- e.g., 128 to match the attention head dimension -- over a reconfigurable routing network.%
A per-PE router (Figure~\ref{fig:elastic_simd2}) toggles two dataflows on a single mode signal: \textit{broadside mode} runs all PEs in parallel across the sequence to stream 1D KV vectors (GEMV) for decode, whereas \textit{cascaded mode} forwards data row-to-row to reuse operands for the fused kernel's thin-GEMMs (prefill), giving systolic-like efficiency without the rigid 2D area overhead.

To avoid spilling intermediate matrices to shared memory for the non-linear operations, each Elastic-SIMD PE couples its multiply-accumulate (MAC) units with an in-situ vector ALU (Figure~\ref{fig:elastic_simd3}): row-wise reduce-max/-sum comparators and an LUT-based exponential unit let the Elastic-SIMD run the entire online-softmax correction within its registers, without stalling the MAC pipeline.

\subsection{NAND-Aware Tiling Algorithm}\label{subsec:tiling}

The fused kernel must be tiled onto the hardware while respecting the physical constraints of NAND flash. Algorithm~\ref{alg:tiling} outlines \MemLLM's NAND-stationary tiling strategy.

\noindent\textbf{Page-aligned K/V tiles.}
To maximize aggregate bandwidth, the KV cache is interleaved across channels at a page granularity.
Since NAND reads whole pages (e.g., 16KB) and, at FP16 precision with head dimension $d=128$, a 16KB page accommodates exactly 32 tokens worth of both K and V vectors, \MemLLM fixes the K/V tile size $T_k$ to 32 tokens.
This ensures that every NAND read perfectly aligns with the page boundary without wasting interface bandwidth.

\begin{figure}[b]
  \vspace{-10pt}
  \centering
  \footnotesize
  \removelatexerror
  \begin{algorithm}[H]
    \setstretch{0.85}
    \SetInd{0.3em}{0.6em}
    \SetAlgoLined
    \caption{NAND-Stationary Tiling \& Fusing}
    \label{alg:tiling}
    \textbf{Input:} $Q_{fused}$ in Global SRAM, KV pages in NAND\\
    \textbf{Output:} $O_{final}$ to Host GPU\\
    \ForEach{NAND Channel $c$ \textbf{in parallel}}{ \label{alg:tiling1}
      \ForEach{KV page $P_{kv}(K_{page}, V_{page})$ in $c$}{ \label{alg:tiling2}
        Load $P_{kv}$ into Local Scratchpad $S_c$\;
        \ForEach{$Q_{tile} \in Q_{fused}$ broadcasted from Global SRAM}{
          // \textit{Phase 1: Output-Stationary} \\ \label{alg:stationary1}
          $A_{tile} \leftarrow Q_{tile} \times K_{page}^T$ \tcp*{Accumulate in Regs}
          // \textit{Phase 2: In-situ Partial Softmax} \\
          $P_{tile}, m_{loc}, \ell_{loc} \leftarrow \text{Vector\_ALU}(A_{tile})$\;
          // \textit{Phase 3: Input-Stationary} \\
          $O_{part} \leftarrow P_{tile} \times V_{page}$ \tcp*{Repurpose Regs}
          Send $(O_{part}, m_{loc}, \ell_{loc})$ to Flash-Coordinator\; \label{alg:stationary2}
        }
      }
    }
    // \textit{Phase 4: Global Reduction (Flash-Coordinator)}\\
    $O_{final} \leftarrow \text{Online\_Softmax\_Reduction}(O_{part}, m_{loc}, \ell_{loc})$\;
    \Return $O_{final}$
  \end{algorithm}
  \vspace{-5pt}
\end{figure}

\noindent\textbf{K-stationary, Q-streaming loop order.}
As shown in Line~\ref{alg:tiling1}--\ref{alg:tiling2}, the loop is structured with KV pages in the outermost loop. Each KV page is fetched from NAND exactly once into the local scratchpad, remaining stationary while $Q$ tiles are streamed through it. Since $Q_{fused}$ is globally broadcasted from the controller's SRAM to all channels simultaneously, the high-bandwidth SRAM interconnect easily hides the broadcast latency behind the slower NAND reads.

\noindent\textbf{Die-level pipelining.}
A NAND page read consists of a sensing phase ($t_R$, 5-20 $\mu$s) and a transfer phase ($t_{Out}$, a few $\mu$s). By striping consecutive token groups across multiple dies in a round-robin fashion, \MemLLM establishes a three-stage pipeline. While die $i$ performs $t_R$, die $i{-}1$ executes $t_{Out}$, and the Elastic-SIMD processes the page from die $i{-}2$. Utilizing double-buffered local scratchpads, this die-level pipelining completely hides the dominant $t_R$ latency.
At the coarser inter-layer granularity, \MemLLM further overlaps the NAND reads for layer $l{+}1$ with the host-side projection and FFN computation of layer $l$, hiding the remaining inter-layer latency across the multi-layer pipeline.

\subsection{DRAM-less Elastic-SIMD Dataflow}\label{subsec:global-accumulator}

The Elastic-SIMD runs each tile through three register-resident phases (Algorithm~\ref{alg:tiling}, Line~\ref{alg:stationary1}--\ref{alg:stationary2}) designed to eliminate intermediate SRAM traffic.
Initially, it operates in an output-stationary mode (Phase~1): it computes $A_{tile} = Q_{tile} \times K_{page}^T$ directly into accumulator registers.
Without spilling $A_{tile}$ to the scratchpad, the hardware transitions to the in-situ partial-softmax phase (Phase~2).
The coupled vector ALU reduces each row to its local maximum, applies the exponential, and reduces again for the sum, turning $A_{tile}$ into the partial probability matrix $P_{tile}$ plus the local max/sum scalars of online attention~\cite{dao2022flashattention,dao2023flashattention}.
Next, the architecture undergoes a dynamic input-stationary transition (Phase~3).
The $P_{tile}$ is repurposed as a fixed input operand within the registers.
The accumulator registers are then reassigned to the output space of $O_{partial}$. As the $V_{page}$ vectors stream from the scratchpad, they are multiplied against the stationary $P_{tile}$.
Whereas a conventional design would write $A_{tile}$ to the scratchpad and read it back, this in-register datapath produces $O_{partial}$ with zero intermediate SRAM writes.
Finally, the Flash-Coordinator (Phase~4) merges the partial outputs via stateful log-sum-exp reduction entirely within its 1MB global SRAM; since $Q_{fused}$ and $O_{final}$ both fit on-chip, the controller's DRAM is bypassed and left undisturbed for FTL operations.

\subsection{MemLLM Software Stack}\label{subsec:software-stack}

\MemLLM is realized through a full-stack system design, which bridges the gap between tensor-level ML frameworks and block-level storage interfaces.
Because conventional POSIX I/O stack operates on logical block addresses and strips the semantics required for tensor offloading, \MemLLM bypasses the virtual file system (VFS) and block layer through the NVMe passthrough interface (\texttt{NVME\_IOCTL\_IO\_CMD}) with a vendor-specific command, \texttt{CMD\_FUSED\_ATTN}.
Note that, since the irregular recomputation mask can exceed several kilobytes for long contexts, it cannot fit within a standard 64-byte submission queue entry (SQE).
The SQE instead carries a lightweight handle to the KV-cache layout together with the physical DMA addresses and lengths of the mask bitvector and $Q_{fused}$, letting the controller fetch every operand by zero-copy peer-to-peer DMA with no OS context switch.
To avoid bounce-buffer copies through host DRAM, \MemLLM uses the Linux \texttt{dma-buf} framework: the kernel driver maps the GPU VRAM holding $Q_{fused}$ into the NVMe driver's DMA mappings, the controller reads it directly into global SRAM over a P2P PCIe transaction, and $O_{final}$ returns along the same path.
Finally, \MemLLM exposes these capabilities transparently to application developers via a PyTorch C++ extension (\texttt{torch.ops.memllm}) that extracts the raw tensor pointers and submits the NVMe commands through Linux \texttt{io\_uring} asynchronous I/O.
Because \texttt{io\_uring} is non-blocking, it lets the host GPU overlap its projection and FFN compute with the storage-side attention, and on the completion interrupt (CQE), it resolves the pending future and returns the completed tensor directly into the PyTorch computation graph.

\section{Evaluation}\label{sec:evaluation}
\subsection{Methodology}\label{subsec:eval_method}

\noindent\textbf{Prototyping and simulation.}
We prototyped the Elastic-SIMD microarchitecture based on Chisel~\cite{dac2016chisel} using the Chipyard~\cite{amid2020chipyard} framework.
For comparison, the reference accelerator designs were prototyped using Hwacha~\cite{lee2015hwacha} for SIMD arrays and Gemmini~\cite{gemmini-dac} for systolic arrays.
All designs were implemented on a 16nm FPGA~\cite{u200} to extract cycle counts, area, and power metrics, ensuring that the distributed Elastic-SIMD clusters and the Global Accumulator operate well within the 3W power envelope of an edge SSD controller.
To estimate the area overhead of \MemLLM, we used CACTI~\cite{muralimanohar2009cacti} and Synopsys Design Compiler at the 45nm process node.
In \MemLLM's per-device flash coordinator, the buffer occupies more than 99\% of the area, accounting for 1.6mm$^2$.
The per-channel accelerator consists of i) the Elastic-SIMD, ii) the Stream Filter, and iii) the scratchpad, occupying 0.11mm$^2$, 0.023mm$^2$, and 0.06mm$^2$, respectively.
Elastic-SIMD is approximately 4.2$\times$ larger than AiF's PE (0.026mm$^2$), as it supports high-bandwidth GEMM and attention operations in addition to GEMV.
Although \MemLLM includes a flash coordinator and a larger PE, its total area is smaller (0.68$\times$) due to its per-channel design compared to per-chip (AiF).

\begin{figure}[b]
  \centering
  \vspace{0pt}
  \begin{minipage}[t]{0.35\linewidth}
    \vspace{0pt}
    \includegraphics[width=1\linewidth]{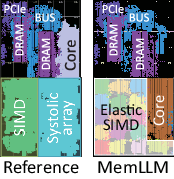}
    \vspace{-20pt}
    \caption{Floorplan.}\label{fig:floorplan}
  \end{minipage}%
  \begin{minipage}[t]{0.02\linewidth}
    \vspace{0pt}
    \hspace{\linewidth}
  \end{minipage}%
  \begin{minipage}[t]{0.64\linewidth}
    \vspace{0pt}
    \small
    \renewcommand*{\arraystretch}{0.3}
    \setlength{\tabcolsep}{0.9pt}
    \resizebox{\linewidth}{!}{%
    \begin{tabular}{@{}cccc@{}}
      \toprule
      \multicolumn{4}{c}{\textbf{HW/SW Configurations}} \\
      \midrule
      \multicolumn{4}{c}{NVIDIA Jetson Orin 16GB~\cite{jetsonorinnx}} \\
      \midrule
      \textbf{Linux kernel}                     & v5.15  & \textbf{CUDA}~\cite{cuda}              & v12.6.10 \\
      \textbf{Pytorch}~\cite{paszke2019pytorch} & v2.8.0 & \textbf{CuDNN}~\cite{chetlur2014cudnn} & v9.3 \\
      \bottomrule
    \end{tabular}
    }
    \resizebox{\linewidth}{!}{%
    \begin{tabular}{@{}cc|cc|cc|cc@{}}
      \toprule
      \multicolumn{8}{c}{\textbf{SimpleSSD Configurations}} \\
      \midrule
      \textbf{\# ch.} & 10 & \textbf{\# way}   & 4 & \textbf{NAND type} & TLC  & \textbf{$t_R$}    & 5-20$\mu$s \\
      \textbf{\# die} & 4  & \textbf{\# plane} & 1 & \textbf{Page size} & 16KB & \textbf{$t_{PR}$} & 500$\mu$s  \\
      \bottomrule
    \end{tabular}
    }
    \vspace{3pt}
    \captionof{table}{Evaluation setup.}\label{tbl:setup}
  \end{minipage}%
  \vspace{-17pt}
\end{figure}

To evaluate end-to-end performance, we developed a comprehensive trace-driven, cycle-level simulator that integrates both host and storage models.
To accurately model host GPU execution time, we natively profiled the LLM inference on an NVIDIA Jetson Orin NX using PyTorch, recording the exact latency of each CUDA kernel.
Operations offloaded to the ISP were simulated using the cycle-accurate performance models derived from our hardware prototypes.
Data transfer latencies between the GPU and the SSD were incorporated by empirically measuring the PCIe interface delays on the host system.
Finally, we integrated SimpleSSD~\cite{jung2017simplessd} to faithfully model the internal storage dynamics, including NAND flash read latencies ($t_R$, $t_{Out}$), die-level pipelining overlap, and channel bandwidth constraints.%
The Jetson Orin NX ships in 8GB and 16GB modules, and we evaluate on the 16GB module.
Because the Orin, like recent edge SoCs and smartphones, adopts a unified memory architecture in which the OS and applications jointly share the physical capacity~\cite{alizadeh2024llm}, dedicating the full 16GB to inference is unrealistic; we therefore conservatively assume that roughly 8GB is available to the RAG workload.
We further analyze the sensitivity to this budget by sweeping the available host memory in Section~\ref{subsec:eval_sensitivity}.

\noindent\textbf{Models and datasets.}
We evaluated the same model and dataset configurations described in Section~\ref{sec:motivation}.
LLM models include OPT~\cite{zhang2022opt} and SmolLM2~\cite{allal2025smollm2smolgoesbig} in FP16 precision, covering both GQA and non-GQA variants (Table~\ref{tbl:models}).
RAG workloads are drawn from eight datasets spanning four domains (Table~\ref{tbl:datasets}).
Selective recomputation follows the algorithms from CacheBlend~\cite{yao2025cacheblend} with a default recomputation ratio of 15\%, and top-$k$=5 retrieval unless stated.

\noindent\textbf{Compared systems.}
We compared \MemLLM against four baselines and evaluated four ablation configurations.
Table~\ref{tbl:compared-systems} summarizes how each system partitions the RAG pipeline, i.e., where it places the model weights, KV cache, and attention computation, together with its CSSD organization and supported operations.
\texttt{GPU-FullCompute} is the baseline.
The GPU performs full prefill over raw context tokens without precomputed KV caches.
KV caches generated during prefill are offloaded to the SSD; the decode phase reads them back over PCIe each iteration.
\texttt{GPU-PreCompute} is the baseline with precomputed KV caches, loaded from the SSD; the GPU then performs selective recomputation.
Both GPU baselines adopt FlexGen-style KV offloading~\cite{pmlr-v202-sheng23a}, staging KV caches in idle host DRAM to avoid redundant SSD round-trips whenever capacity permits, and prefetching them to overlap SSD-to-host transfers with computation.
As shown in Section~\ref{subsec:motiv_overall}, the I/O overhead of loading precomputed KV caches can negate the computational savings on-device.
\texttt{InstAttention}~\cite{pan2025instattention} adopts a decode-only ISP that offloads decode-phase attention to the storage device, eliminating decode-time KV transfers over PCIe.
However, prefill (including recomputation) remains on the host GPU, triggering a massive PCIe round-trip of the KV cache.
\texttt{AiF}~\cite{lee2025aif} is an in-flash processing architecture that places model weight computation in the storage device while performing attention on the host GPU.
Because the KV cache resides in GPU memory, it achieves strong performance when contexts fit in memory, but encounters fatal out-of-memory (OOM) errors for long contexts.

\begin{table}[b]
  \centering
  \small
  \renewcommand*{\arraystretch}{1.0}
  \setlength{\tabcolsep}{3pt}
  \resizebox{\linewidth}{!}{%
  \begin{tabular}{@{}lccccc@{}}
    \toprule
    \multicolumn{1}{c}{\multirow{2}{*}{\textbf{\shortstack{Arch.
    \\ Name}}}} & \multirow{2}{*}{\textbf{\shortstack{Model \\ Weight}}} &\multirow{2}{*}{\textbf{KV\$}} & \multirow{2}{*}{\textbf{\shortstack{Attention \\ Compute}}} & \multirow{2}{*}{\textbf{\shortstack{CSSD \\ Type}}} & \multirow{2}{*}{\textbf{Supported ops.}} \\
                                                       &     &     &     &            &  \\
    \midrule
    \textbf{Baseline}                            & GPU  & SSD  & GPU  & -          & - \\
    \textbf{InstAtt}\cite{pan2025instattention} & GPU  & CSSD & CSSD & Per-device & Sparse attention \\
    \textbf{AiF}\cite{lee2025aif}               & CSSD & GPU  & GPU  & Per-chip   & GEMV \\
    \midrule
    \multirow{2}{*}{\textbf{\shortstack{MemLLM \\ (Ours)}}} & \multirow{2}{*}{GPU} & \multirow{2}{*}{CSSD} & \multirow{2}{*}{CSSD} & \multirow{2}{*}{Per-channel}  & \multirow{2}{*}{\shortstack{GEMV/GEMM \\ Dense/Sparse attn.}} \\
    &     &     &     &            &  \\
    \bottomrule
  \end{tabular}
  }
  \vspace{3pt}
  \caption{Comparison of compared systems.}\label{tbl:compared-systems}
  \vspace{-17pt}
\end{table}

For \MemLLM, we evaluate the full system and three intermediate ablation configurations to isolate the contribution of each design choice.
\texttt{MemLLM-Base} adopts a single, centralized Elastic-SIMD accelerator within the SSD controller that performs end-to-end RAG attention.
KV caches are read from NAND through the internal SSD bus.
\texttt{MemLLM-Fused} adds the fused recomputation-query kernel (Section~\ref{subsec:fused-kernel}), halving the NAND read volume by processing recomputation and query attention in a single pass.
\texttt{MemLLM-PerCh} further distributes the Elastic-SIMD across per-channel endpoints (Section~\ref{subsec:elastic_simd}), eliminating the internal bus bottleneck and exploiting aggregate NAND channel bandwidth.
The total compute budget ($\sim$3W) is preserved.
Finally, \texttt{MemLLM} adds the Flash-Coordinator with the global SRAM accumulator and stateful reduction engine (Section~\ref{subsec:global-accumulator}), enabling DRAM-less operation and completely eliminating PCIe egress bottlenecks.

\begin{figure}
  \centering
  \vspace{0pt}
  \includegraphics[width=1\linewidth]{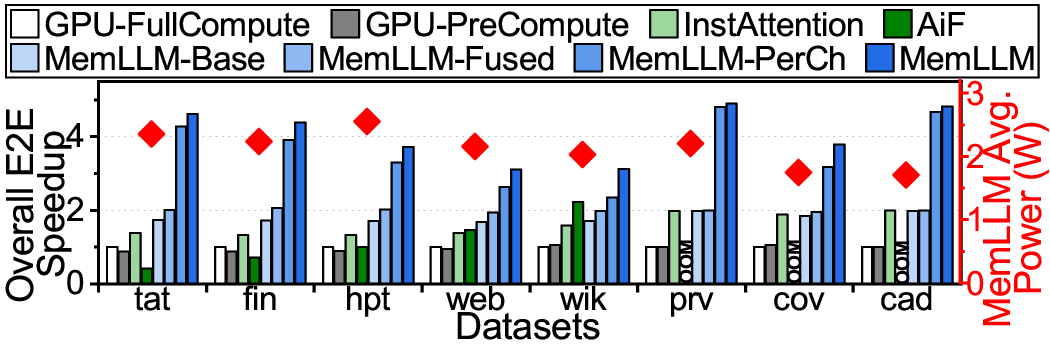}
  \vspace{-13pt}
  \caption{End-to-end speedup and average power.}\label{fig:eval-overall}
  \vspace{-10pt}
\end{figure}

\subsection{Overall Performance}\label{subsec:eval_overall}

\noindent\textbf{Overall speedup.}
Figure~\ref{fig:eval-overall} presents the end-to-end RAG inference speedup of all systems, normalized to the \texttt{GPU-FullCompute} baseline, across eight datasets.
Overall, \texttt{MemLLM} achieves an average speedup of 4.06$\times$ over the GPU baseline and 4.18$\times$ over \texttt{GPU-PreCompute}, outperforming all compared systems across every dataset.
The datasets were sorted in order of the average context length.
\texttt{GPU-PreCompute} performs slightly worse than the baseline as the context size increases, exhibiting an average performance drop of 2.86\%.
This confirms the I/O wall identified in Section~\ref{subsec:motiv_overall}: the massive PCIe transfer overhead of loading the precomputed KV caches outweighs the computational savings of prefill offloading.

Both \texttt{InstAttention} and \texttt{AiF} leverage ISP designs to minimize KV cache movement during the decode phase, achieving average speedups of 1.61$\times$ and 1.17$\times$, respectively.
Notably, both systems still require the precomputed KV cache to be transferred to the host for prefill, meaning they remain fundamentally bottlenecked by the PCIe round-trip overhead required for selective recomputation.
Beyond this shared prefill bottleneck, they exhibit distinct limitations during the decode phase due to their fundamentally opposite offloading strategies.
\texttt{InstAttention} offloads the KV cache to the ISP while keeping model weights in the host GPU.
While this enables the system to handle long contexts without memory capacity issues, its decode performance is ultimately restricted; as the context grows, streaming the massive KV cache internally within the SSD becomes bottlenecked by the device's internal system bus and DRAM bandwidth.
Conversely, \texttt{AiF} offloads model weights to the ISP while retaining the KV cache in GPU memory.
This allows \texttt{AiF} to exploit high-bandwidth VRAM reads for the KV cache during each decode iteration.
Consequently, as the context length increases, \texttt{AiF} delivers significantly higher speedups compared to \texttt{InstAttention}.
However, \texttt{AiF} is strictly bounded by the GPU's memory capacity, resulting in fatal Out-Of-Memory (OOM) errors for datasets with longer contexts (e.g., \texttt{prv}, \texttt{cov}, \texttt{cad}).

\MemLLM systematically resolves these bottlenecks through its cumulative I/O minimization design.
\texttt{MemLLM-Base} eliminates the precomputed KV cache PCIe transfers entirely by executing the selective recomputation directly within the ISP.
This provides an average speedup of 1.80$\times$ compared to baseline.
\texttt{MemLLM-Fused} utilizes the fused kernel to effectively halve the NAND access volume, yielding an additional 1.11$\times$ speedup, on average.
\texttt{MemLLM-PerCh} exploits the aggregate NAND channel bandwidth thanks to its distributed per-channel accelerators, providing an additional 1.82$\times$ speedup, on average.
Finally, the full \texttt{MemLLM} incorporates a global reduction logic, preventing intermediate partial outputs ($O_{partial}$) from spilling into the storage's internal DRAM.
This DRAM-less execution provides the final performance leap of 1.12$\times$, on average, maximizing the end-to-end speedup.

\begin{figure}
  \centering
  \vspace{0pt}
  \includegraphics[width=1\linewidth]{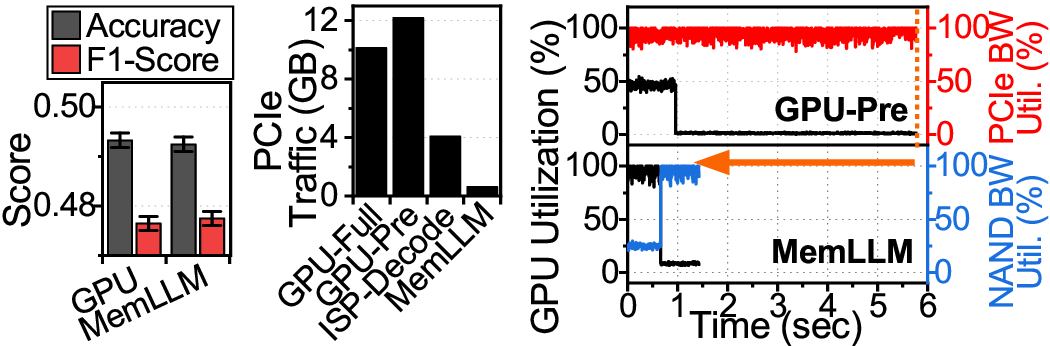}
  \begin{minipage}[t]{0.165\linewidth}
    \vspace{0pt}
    \vspace{-12pt}
    \caption{Accuracy.}\label{fig:eval-accuracy}
  \end{minipage}%
  \begin{minipage}[t]{0.1\linewidth}
    \hfill
  \end{minipage}%
  \begin{minipage}[t]{0.25\linewidth}
    \vspace{0pt}
    \vspace{-12pt}
    \caption{PCIe traffic.}\label{fig:eval-pcie}
  \end{minipage}%
  \begin{minipage}[t]{0.05\linewidth}
    \hfill
  \end{minipage}%
  \begin{minipage}[t]{0.4\linewidth}
    \vspace{-7pt}
    \caption{Timeseries.}\label{fig:eval-dram-timeseries}
  \end{minipage}%
  \begin{minipage}[t]{0.05\linewidth}
    \hfill
  \end{minipage}%
  \vspace{-15pt}
\end{figure}

\noindent\textbf{Average power.}
The right y-axis of Figure~\ref{fig:eval-overall} shows the average power consumption of \MemLLM for each dataset.
Based on the operating power measured on the FPGA, \MemLLM consumes 2.12W on average and 2.55W at peak, well within the 3W power budget.
As the power differs by up to 1.9$\times$ between the compute-intensive fused recomputation kernel and the memory-intensive decode kernel, the average power varies in proportion to the prefill-to-decode time ratios among datasets.
Since the large parasitic capacitances and constant leakage of the programmable FPGA fabric are absent from the fixed wiring and dedicated logic of an ASIC, the deployed power is expected to drop well below this measurement~\cite{kuon2007measuring}.

\noindent\textbf{Accuracy.}
As discussed earlier, \MemLLM performs mathematically the same computation as CacheBlend, differing only in the order of operations (Section~\ref{subsec:fused-kernel}).
Figure~\ref{fig:eval-accuracy} reports the downstream task (TriviaQA) accuracy and F1-score of the baseline and \MemLLM.
The results show that \MemLLM exhibits a difference of less than 0.3\% on both metrics, within the margin of error.
In terms of F1-score, this is 1.35$\times$ smaller than the 0.002 deviation reported by CacheBlend.

\noindent\textbf{PCIe traffic reduction.}
Figure~\ref{fig:eval-pcie} quantifies the total PCIe data transferred during a single RAG inference.
Both \texttt{GPU-Full} and \texttt{GPU-Pre} suffer massive decode-phase PCIe traffic from continuously offloading and reloading the KV cache, and \texttt{GPU-Pre} aggravates this by additionally loading the entire precomputed KV cache from storage during prefill.
Both \texttt{ISP-Decode} architectures (\texttt{InstAttention} and \texttt{AiF}) eliminate the decode-phase KV cache movement, but since they cannot process the sparse recomputation internally, they stream the precomputed KV cache across PCIe twice during prefill: once for selective recomputation and again for query attention.
\texttt{MemLLM} virtually eliminates this bottleneck.
To generate the initial recomputation mask at the host, it only fetches the first layer's precomputed KV cache ($<$5\% of the total volume); thereafter, the only data crossing PCIe are the lightweight query vectors ($q_{new}$) and final attention outputs ($O_{final}$) of each decode step.
The final PCIe footprint is near-zero, confirming that \MemLLM decouples the PCIe bandwidth requirement from context length.

\noindent\textbf{Utilization.}
Figure~\ref{fig:eval-dram-timeseries} shows the GPU and PCIe/NAND bandwidth utilization of \texttt{GPU-Pre} and \texttt{MemLLM} over time.
\texttt{GPU-Pre} aggressively prefetches the KV cache from storage and saturates nearly 100\% of the PCIe bandwidth.
However, even with the PCIe link fully utilized, this bandwidth is insufficient to keep the GPU fed, so the GPU idles waiting for the data movement, its utilization dropping close to 0\%.
In contrast, \texttt{MemLLM} eliminates the data movement over PCIe and saturates aggregated NAND bandwidth inside the SSD, completing the same task approximately 4$\times$ faster.
During the initial fused HKVD recomputation, \texttt{MemLLM} incurs NAND write I/O for the newly computed KV caches.
However, during the recomputation phase, \texttt{MemLLM}'s GPU utilization reaches 100\%.
This indicates that even the low write bandwidth of NAND is sufficient to keep the GPU from stalling.

\begin{figure}
  \centering
  \vspace{0pt}
  \includegraphics[width=1\linewidth]{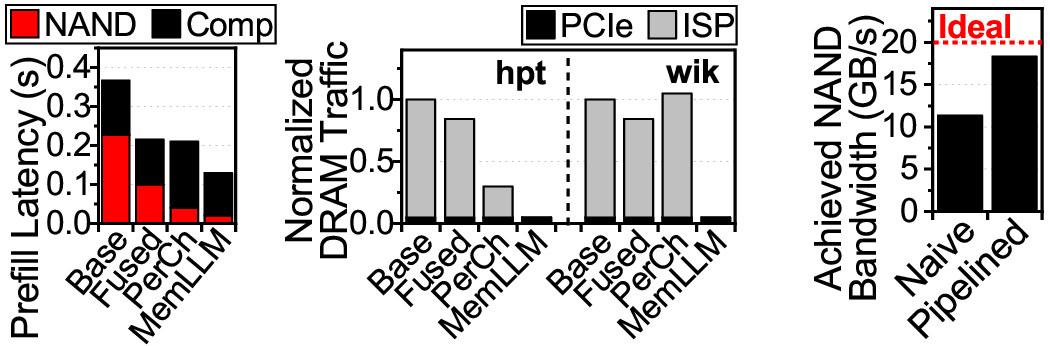}
  \begin{minipage}[t]{1\linewidth}
    \vspace{2pt}
    \caption{Ablation analysis.}\label{fig:eval-ablation}
    \vspace{-30pt}
    \begin{subfigure}{1\linewidth}
      \renewcommand*{\arraystretch}{0.3}
      \begin{tabular}{
        p{\dimexpr.345\linewidth-1.33\tabcolsep}
        p{\dimexpr.295\linewidth-1.33\tabcolsep}
        p{\dimexpr.355\linewidth-1.33\tabcolsep}
        }
        \caption{Prefill breakdown.}\label{fig:eval-ablation-prefill} &
        \caption{DRAM traffic.}\label{fig:eval-ablation-dram} &
        \caption{NAND Pipelining.}\label{fig:eval-pipelining}
      \end{tabular}
    \end{subfigure}
    \vspace{-15pt}
  \end{minipage}%
  \vspace{0pt}
\end{figure}

\subsection{Ablation Analysis}

\noindent\textbf{Prefill latency breakdown.}
Figure~\ref{fig:eval-ablation-prefill} decomposes the prefill latency of each \MemLLM configuration into exposed NAND read time and computation time, where the latter includes the data movement overhead of accessing the SSD's internal DRAM.
\texttt{MemLLM-Base} is heavily bottlenecked by the exposed NAND latency.
The fused kernel (\texttt{MemLLM-Fused}) halves the NAND read time by co-processing selective recomputation and query attention in a single pass, while computation time stays roughly constant.
Distributing to the channel endpoints (\texttt{MemLLM-PerCh}) exploits the aggregate NAND channel bandwidth and dramatically collapses the exposed NAND time, but reveals a bottleneck shift: as the concurrent per-channel accelerators simultaneously stream their partial outputs ($O_{partial}$) to the controller, they flood the SSD's limited internal DRAM, and the resulting read-modify-write contention inflates the compute latency; showing that parallelizing the compute fabric alone is insufficient without redesigning the reduction dataflow.
The full \texttt{MemLLM} resolves this with its DRAM-less design: accumulating the partial outputs in-situ in the global SRAM reduction logic bypasses the DRAM and restores optimal latency, while die-level pipelining overlaps the remaining NAND reads with computation to hide the NAND latency almost completely (evaluated in detail shortly).

\noindent\textbf{DRAM transfer analysis.}
Figure~\ref{fig:eval-ablation-dram} presents the normalized DRAM traffic within the SSD controller, decomposed into PCIe transfer buffering and ISP computation overhead.
We compared \texttt{hpt} and \texttt{wik} as representative short- and long-context datasets.
In \texttt{MemLLM-Base}, the centralized Elastic-SIMD must read the entire precomputed KV cache from NAND into internal DRAM before processing, generating massive traffic.
The fused kernel (\texttt{MemLLM-Fused}) reduces NAND read passes, but intermediate results still thrash the DRAM, keeping the I/O footprint high.
Distributing compute to the channels (\texttt{MemLLM-PerCh}) offloads KV cache processing to the endpoints, eliminating central KV staging, but the resulting DRAM traffic diverges drastically with context length.
For short contexts (\texttt{hpt}), the $K$ matrix needs fewer tiling iterations, producing few partial outputs ($O_{partial}$) and thus a significant traffic drop.
However, for long contexts (\texttt{wik}), extensive $K$ tiling floods all channels with $O_{partial}$ vectors, and the repeated read-modify-write required to accumulate them explodes the controller's DRAM traffic, even surpassing \texttt{Base}.
The full \texttt{MemLLM} resolves this: the central Flash-Coordinator processes the $O_{partial}$ vectors in-situ in a global SRAM buffer, and this DRAM-less accumulation eradicates the computation-related traffic, leaving only the minimal baseline PCIe transfer I/O (for $Q_{fused}$ and $O_{final}$).

\begin{figure}
  \centering
  \vspace{0pt}
  \includegraphics[width=1\linewidth]{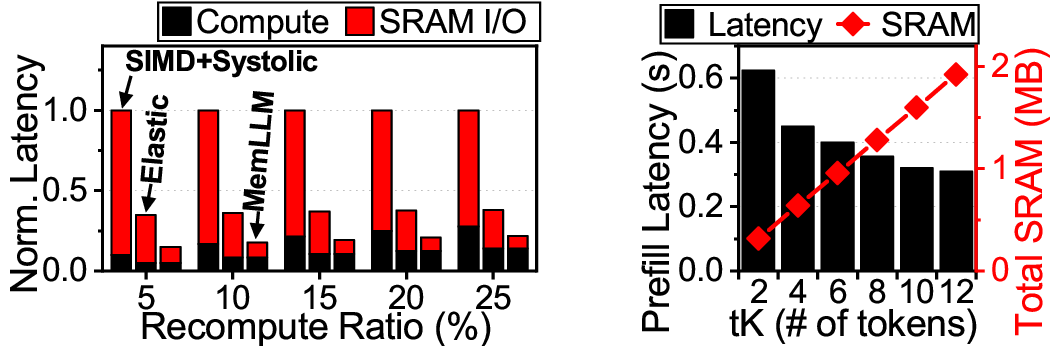}
  \vspace{-4pt}
  \caption{Sensitivity analysis.}\label{fig:eval-sensitivity}
  \vspace{-30pt}
  \begin{subfigure}{1\linewidth}
    \renewcommand*{\arraystretch}{0.3}
    \begin{tabular}{
      p{\dimexpr.60\linewidth-1.33\tabcolsep}
      p{\dimexpr.34\linewidth-1.33\tabcolsep}
      }
      \caption{Recompute ratio.}\label{fig:eval-sensitivity-recompute} &
      \caption{Tile size.}\label{fig:eval-sensitivity-tiling}
    \end{tabular}
  \end{subfigure}
  \vspace{-10pt}
\end{figure}

\noindent\textbf{Pipeline efficiency.}
Figure~\ref{fig:eval-pipelining} quantifies die-level pipelining in the distributed Elastic-SIMD accelerators, comparing the achieved aggregate NAND bandwidth between a naive sequential execution and the proposed three-stage pipelined execution using double-buffered local scratchpads.
Under the naive approach, the NAND sensing time ($t_R$) is exposed to the critical path, frequently stalling the compute units and dropping the bandwidth to nearly half the theoretical maximum ($\sim$11 GB/s).
The pipelined execution overlaps NAND latencies with computation of previously fetched KV pages; this double-buffering completely hides the read latencies behind the compute, pushing effective throughput to $\sim$19 GB/s.

\subsection{Sensitivity and Scalability Analysis}\label{subsec:eval_sensitivity}

\noindent\textbf{Recomputation ratio.}
Figure~\ref{fig:eval-sensitivity-recompute} evaluates the impact of the recomputation ratio (5\%--25\%) on compute and SRAM I/O latency across three architectures under a fixed power budget.
The decoupled \texttt{SIMD+Systolic} baseline suffers from massive SRAM I/O to exchange intermediate results between disparate compute units, alongside degraded effective FLOPS due to resource division.
\texttt{Elastic} drastically reduces this SRAM traffic via register-centric stationary dataflows and in-situ operations.
However, applying the recomputed tokens from the host to the precomputed KV cache still incurs SRAM staging overhead.
\MemLLM's full design (\texttt{Elastic-Filter}) resolves this by utilizing the hardware Stream Filter to fuse recomputed KV entries on-the-fly during NAND loading.
This completely eliminates the remaining SRAM bottleneck, ensuring minimal latency and robust efficiency across a wide range of recomputation ratios, including the 18\% suggested by CacheBlend~\cite{yao2025cacheblend}.

\begin{figure}
  \centering
  \vspace{0pt}
  \includegraphics[width=1\linewidth]{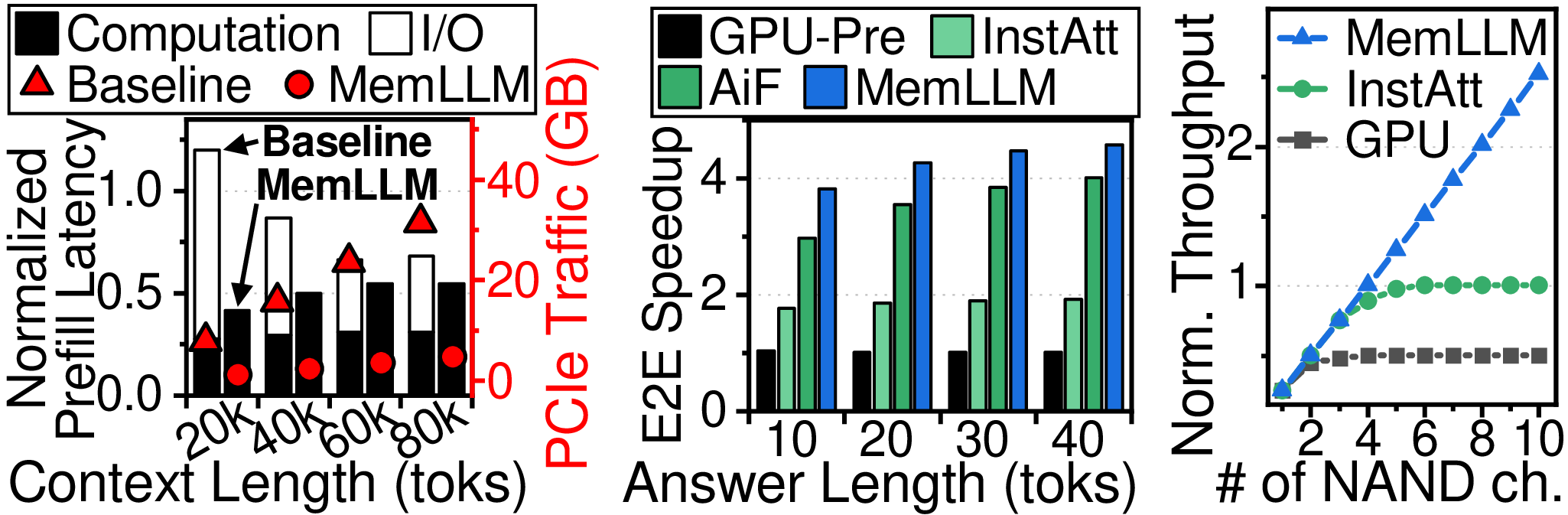}
  \vspace{-3pt}
  \caption{Scalability analysis.}\label{fig:eval-scalability}
  \vspace{-32pt}
  \begin{subfigure}{1\linewidth}
    \renewcommand*{\arraystretch}{0.3}
    \begin{tabular}{
      p{\dimexpr.31\linewidth-1.33\tabcolsep}
      p{\dimexpr.02\linewidth-1.33\tabcolsep}
      p{\dimexpr.3\linewidth-1.33\tabcolsep}
      p{\dimexpr.33\linewidth-1.33\tabcolsep}
      }
      \caption{Context length.}\label{fig:eval-scalability-ctx} & &
      \caption{Answer length.}\label{fig:eval-scalability-ans} &
      \caption{NAND channels.}\label{fig:eval-scalability-ch}
    \end{tabular}
  \end{subfigure}
  \vspace{-5pt}
\end{figure}

\noindent\textbf{Tile size.}
Figure~\ref{fig:eval-sensitivity-tiling} sweeps the KV tile size ($t_K$) to evaluate its impact on prefill latency and total SRAM capacity.
Increasing $t_K$ lets the Elastic-SIMD process larger compute batches at once, better hiding the exposed NAND read latency and reducing prefill time, but the required SRAM scales linearly with it.
The latency improvement saturates beyond $t_K=12$, where the total SRAM footprint (aggregated across all channels) reaches approximately 2MB.

\noindent\textbf{Context length.}
Figure~\ref{fig:eval-scalability-ctx} evaluates prefill latency and PCIe traffic as context length increases.
As context grows, the number of tokens requiring recomputation rises proportionally, yielding a denser GEMM workload that the baseline runs faster on its high-performance host GPU than \MemLLM's power-constrained storage accelerators.
However, this compute advantage is entirely eclipsed by I/O: as shown on the right y-axis, the baseline's PCIe traffic scales linearly with context length, whereas \MemLLM confines all KV cache processing within storage, maintaining a strict $O(1)$ PCIe footprint regardless of context size.
Consequently, while \MemLLM's compute time scales with the workload, eliminating the I/O wall lets it support the maximum context windows of modern LLMs without being bottlenecked.

\begin{figure}
  \centering
  \vspace{0pt}
  \includegraphics[width=1\linewidth]{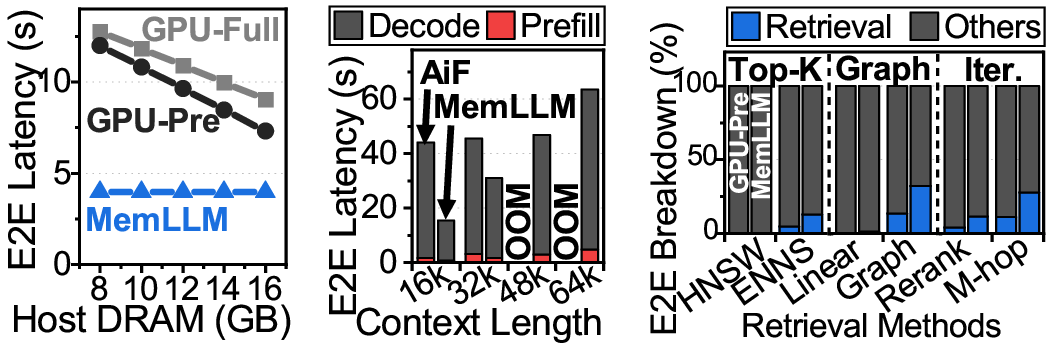}
  \begin{minipage}[t]{0.27\linewidth}
    \vspace{-10pt}
    \caption{Memory capacity.}\label{fig:eval-dram-sensitivity}
  \end{minipage}%
  \begin{minipage}[t]{0.035\linewidth}
    \hfill
  \end{minipage}%
  \begin{minipage}[t]{0.28\linewidth}
    \vspace{-10pt}
    \caption{Long-answer scenario.}\label{fig:eval-ctx-ret-long}
  \end{minipage}%
  \begin{minipage}[t]{0.045\linewidth}
    \hfill
  \end{minipage}%
  \begin{minipage}[t]{0.36\linewidth}
    \vspace{-10pt}
    \caption{Retrieval methods.}\label{fig:eval-ctx-ret-ret}
  \end{minipage}%
  \vspace{-20pt}
\end{figure}

\noindent\textbf{Answer length.}
Figure~\ref{fig:eval-scalability-ans} evaluates end-to-end speedup as the generated answer length scales, where the decode phase takes a larger fraction of inference time.
Both \texttt{InstAttention} and \texttt{AiF} eliminate decode-phase PCIe transfers through their ISP designs, yielding significant speedups over the \texttt{GPU-Pre} baseline; \texttt{AiF} scales especially well here by retaining the KV cache in high-bandwidth VRAM to minimize per-token decode latency.
However, \texttt{MemLLM} continues to dominate: although it must stream the KV cache from NAND flash during decode, its distributed per-channel Elastic-SIMD accelerators meet the throughput requirement by exploiting aggregate NAND channel bandwidth.
Compounded with its prefill-phase savings, \texttt{MemLLM} sustains the highest end-to-end speedup across all evaluated answer lengths.
We further evaluate longer-answer scenarios in Figure~\ref{fig:eval-ctx-ret-long}.

\noindent\textbf{Channel count.}
Figure~\ref{fig:eval-scalability-ch} evaluates normalized decode throughput as the number of NAND channels increases.
The \texttt{GPU} baseline stays flat regardless of storage parallelism, as its performance is throttled by the external PCIe bandwidth.
\texttt{InstAtt} bypasses this PCIe bottleneck during decode but saturates quickly, as its centralized ISP design shifts the bottleneck to the SSD's internal system bus.
In contrast, \texttt{MemLLM} scales near-linearly: by distributing Elastic-SIMD compute units directly to the channel endpoints, it utilizes the aggregate NAND bandwidth without internal contention.

\noindent\textbf{Host DRAM scalability.}
Figure~\ref{fig:eval-dram-sensitivity} reports the end-to-end latency as the host DRAM capacity is increased from 8GB to 16GB.
Even at 16GB of host memory, \texttt{\MemLLM} maintains a 1.83$\times$ performance advantage over the baseline.
Since \texttt{\MemLLM} keeps its host memory usage to a minimum, its latency remains constant.
In contrast, the \texttt{GPU-Full} and \texttt{GPU-Pre} baselines improve progressively with host memory capacity, as their FlexGen-style offloading keeps a larger fraction of the KV cache GPU-resident and thereby moves less over PCIe per iteration.
Under the realistic shared-memory budget established in Section~\ref{subsec:eval_method}, the KV cache cannot fully reside in memory for long contexts, so PCIe transfers remain unavoidable for the baselines.

\noindent\textbf{Long-answer scalability.}
Figure~\ref{fig:eval-ctx-ret-long} extends the answer-length analysis (Figure~\ref{fig:eval-scalability-ans}) to a long-answer regime, sweeping the context length with the answer length fixed at 100 tokens.
Here the decode phase dominates, so AiF's high-bandwidth VRAM-resident KV cache lets it degrade more gracefully than \MemLLM as the context grows.
This advantage, however, comes with a capacity trade-off: since AiF stages the entire KV cache in GPU memory, the processable context is bounded to below 48k, whereas the crossover point would only arise at still larger contexts.
Across the range that both systems can serve, \MemLLM retains its advantage, while supporting the longer contexts that exceed AiF's memory budget.

\noindent\textbf{Retrieval methods.}
To demonstrate that \MemLLM is not confined to a single top-$k$ similarity search, we evaluate three classes of retrieval methods in Figure~\ref{fig:eval-ctx-ret-ret}, decomposing the end-to-end latency into retrieval and others: i) top-$k$ similarity search (HNSW, ENNS)~\cite{malkov2020efficient,douze2024faiss}, ii) graph-based retrieval (linear scan, graph traversal)~\cite{edge2024local,zhuang2025linearrag}, and iii) iterative retrieval (reranking, multi-hop)~\cite{nogueira2020document,trivedi2023interleaving}.
In iterative methods such as reranking and multi-hop, the final chunk is determined only after repeated LLM-based rescoring or multi-stage querying.
Unlike a single top-$k$ search, these methods introduce additional latency as a separate front-end stage.
However, once the final chunk is fixed, the subsequent attention pipeline is identical regardless of the retrieval method, so \MemLLM accelerates this dominant segment.
In practice, retrieval accounts for at most 30\% of the end-to-end latency even in the heaviest graph and multi-hop cases.
By accelerating the remaining dominant segment, \MemLLM sustains its gains over the baseline across all retrieval methods.

\begin{figure}
  \centering
  \vspace{0pt}
  \includegraphics[width=1\linewidth]{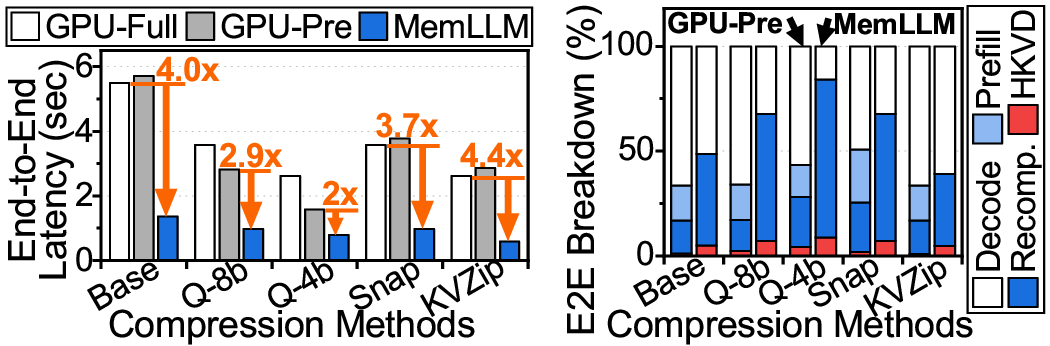}
  \vspace{-2pt}
  \caption{KV compression methods.}\label{fig:eval-compression}
  \vspace{-31pt}
  \begin{subfigure}{1\linewidth}
    \renewcommand*{\arraystretch}{0.3}
    \begin{tabular}{
      p{\dimexpr.55\linewidth-1.33\tabcolsep}
      p{\dimexpr.45\linewidth-1.33\tabcolsep}
      }
      \caption{End-to-end latency.}\label{fig:eval-compression-latency} &
      \caption{Latency breakdown.}\label{fig:eval-compression-breakdown}
    \end{tabular}
  \end{subfigure}
  \vspace{-5pt}
\end{figure}

\noindent\textbf{KV compression methods.}
Figure~\ref{fig:eval-compression-latency} evaluates the baseline and \MemLLM under optimizations that directly reduce the KV cache size: quantization to reduce the per-token size (\texttt{Q-8b}, \texttt{Q-4b}) and entry-count reduction via \texttt{SnapKV}~\cite{li2024snapkv} and \texttt{KVZip}~\cite{kim2026kvzip}, representative runtime- and static-based methods, respectively.
\texttt{SnapKV} can compress only after the query prefill completes, whereas \texttt{KVZip} compresses the contexts through preprocessing.
By shrinking the KV cache, these methods ease the limited PCIe bandwidth that bottlenecks edge environments; however they likewise cut the NAND reads of \MemLLM, so \MemLLM sustains a comparable 2.0--4.4$\times$ end-to-end speedup over the baseline across all methods.
Crucially, these techniques are orthogonal to and composable with \MemLLM: while they shrink the I/O wall and thus narrow \MemLLM's margin, they do not remove it, since even compressed caches can exceed on-device memory for long contexts and still cross PCIe.
Figure~\ref{fig:eval-compression-breakdown} breaks the end-to-end latency into HKVD identification, recomputation, query prefill, and the decode steps.
HKVD identification and \MemLLM's recomputation are both compute-bound, dominated respectively by the host GPU's full-attention pass and the fused kernel's prefill operations.
They are thus barely affected by compression, and their share grows with the compression ratio, reaching 8.8\% and 75.5\% under \texttt{Q-4b}.
Since \texttt{SnapKV} compresses only after prefill, it benefits only the decode phase, whereas \texttt{KVZip} benefits all phases alike; consequently \texttt{KVZip}'s recomputation share is 26.2\%p lower than \texttt{SnapKV}'s.
Even under aggressive compression, \MemLLM retains an advantage, as the recomputation still reads the precomputed KV that the baseline must stream across PCIe but \MemLLM keeps within storage.

\section{Discussion}\label{sec:discussion}
\noindent\textbf{Scope.}
\MemLLM targets on-device RAG with precomputed document KV caches, selective recomputation, long contexts, and short answers; the regime where the precompute-then-recompute prefill and its KV-cache I/O dominate.
Its benefit narrows outside this regime: long answers shift the bottleneck to decode (Figure~\ref{fig:eval-ctx-ret-long}), and a memory-resident KV cache narrows the I/O wall, though \MemLLM still leads (Figure~\ref{fig:eval-dram-sensitivity}).
As Section~\ref{sec:motivation} shows, this memory-constrained, long-context regime is the common case for practical on-device RAG.

\noindent\textbf{Deployment requirements.} \MemLLM assumes a computational storage device that integrates per-channel accelerators with a central coordination logic.
Although \MemLLM adopts a custom NVMe command, it i) occupies a vendor-specific opcode reserved by the specification, and ii) requires no host-side kernel modification when issued via the io\_uring-based NVMe passthrough introduced in Linux 5.19.
The GPUDirect RDMA needed for GPU--device P2P DMA is already supported on the Jetson Orin; unified-memory SoCs like modern smartphones, where the GPU and storage share one physical address space, further simplify this path relative to conventional systems. NVIDIA's GPUDirect Storage does not yet cover such SoCs, but its support is expanding rapidly.
Once these conditions are met, our software stack handles the GPU/SSD split execution and exposes it as a single PyTorch operation.
As demand for RAG over sensitive personal data spreads from medical and legal PCs toward smartphones, \MemLLM could enable efficient on-device RAG on such devices.

\section{Related Works}\label{sec:related}
\noindent\textbf{Precomputed KV cache.}
Prompt caching~\cite{gim2024prompt,schroeder2025vcache,zhu2024efficientpromptcachingembedding,sharma2013greencache} reuses the KV caches of frequent system prompts or similar queries to skip redundant prefill.
For RAG, combining independently precomputed document caches disrupts cross-attention, so CacheBlend~\cite{yao2025cacheblend} and CacheClip~\cite{yang2025cacheclip} apply selective recomputation, updating only a minimal token subset to restore quality.
These software optimizations speed up high-bandwidth servers but hit an I/O wall on resource-constrained edge devices; \MemLLM is the first to mitigate it.

\noindent\textbf{ISP for LLMs.}
Recent ISP studies alleviate the PCIe bottleneck of offloading-based LLM inference.
Decode-phase offloading~\cite{pan2025instattention,jang2026hilos} pushes memory-intensive attention to the SSD to eliminate KV cache transfers during decode; AiF~\cite{lee2025aif} moves weight computation into flash to free GPU VRAM for the KV cache; and RAGX~\cite{mahapatra2025storage} accelerates RAG's vector-database retrieval within storage.
While effective for standard LLMs, these are mismatched for on-device RAG over precomputed KV caches: decode-only ISPs cannot handle the compute-intensive selective recomputation, and weight-offloading designs trigger OOM for long contexts.
\MemLLM is the first to support end-to-end RAG compute.

\section{Conclusion}\label{sec:conclusion}
\MemLLM is an end-to-end in-storage processing architecture that executes the complete RAG attention pipeline directly within the SSD.
By leveraging distributed, per-channel Elastic-SIMD accelerators, \MemLLM fully exploits the aggregate NAND channel bandwidth and shifts all KV cache-related computations to the storage side to minimize PCIe transfers.
Furthermore, \MemLLM mitigates internal I/O overheads through the co-design of a fused recomputation-query kernel and a DRAM-less global reduction engine.
Evaluated across eight real-world RAG datasets and six on-device LLM models, \MemLLM achieves an average end-to-end speedup of 4.06$\times$ compared to GPU-based systems.

\section*{Acknowledgment}
The authors thank anonymous reviewers of ASPLOS'26 and MICRO'26 for their constructive feedback.
This work is protected by one or more patents.
Myoungsoo Jung is the corresponding author (mj@camelab.org).

\bibliographystyle{IEEEtran}
\bibliography{ref.bib}

@article{amid2020chipyard,
  author   = {Amid, Alon and Biancolin, David and Gonzalez, Abraham and Grubb, Daniel and Karandikar, Sagar and Liew, Harrison and Magyar, Albert and Mao, Howard and Ou, Albert and Pemberton, Nathan and Rigge, Paul and Schmidt, Colin and Wright, John and Zhao, Jerry and Shao, Yakun Sophia and Asanović, Krste and Nikolić, Borivoje},
  journal  = {IEEE Micro},
  title    = {Chipyard: Integrated Design, Simulation, and Implementation Framework for Custom SoCs},
  year     = {2020},
  volume   = {40},
  number   = {4},
  pages    = {10-21},
  doi      = {10.1109/MM.2020.2996616}
}

@article{yang2018hotpotqa,
  title   = {HotpotQA: A dataset for diverse, explainable multi-hop question answering},
  author  = {Yang, Zhilin and Qi, Peng and Zhang, Saizheng and Bengio, Yoshua and Cohen, William W and Salakhutdinov, Ruslan and Manning, Christopher D},
  journal = {arXiv preprint arXiv:1809.09600},
  year    = {2018}
}

@article{joshi2017triviaqa,
  title   = {Triviaqa: A large scale distantly supervised challenge dataset for reading comprehension},
  author  = {Joshi, Mandar and Choi, Eunsol and Weld, Daniel S and Zettlemoyer, Luke},
  journal = {arXiv preprint arXiv:1705.03551},
  year    = {2017}
}

@inproceedings{moller2020covid,
  title     = {COVID-QA: A question answering dataset for COVID-19},
  author    = {M{\"o}ller, Timo and Reina, Anthony and Jayakumar, Raghavan and Pietsch, Malte},
  booktitle = {Proceedings of the 1st Workshop on NLP for COVID-19 at ACL 2020},
  year      = {2020}
}

@article{jin2019pubmedqa,
  title   = {Pubmedqa: A dataset for biomedical research question answering},
  author  = {Jin, Qiao and Dhingra, Bhuwan and Liu, Zhengping and Cohen, William W and Lu, Xinghua},
  journal = {arXiv preprint arXiv:1909.06146},
  year    = {2019}
}

@article{hendrycks2021cuad,
  title   = {Cuad: An expert-annotated nlp dataset for legal contract review},
  author  = {Hendrycks, Dan and Burns, Collin and Chen, Anya and Ball, Spencer},
  journal = {arXiv preprint arXiv:2103.06268},
  year    = {2021}
}

@article{chen2021finqa,
  title     = {{F}in{QA}: A Dataset of Numerical Reasoning over Financial Data},
  author    = {Chen, Zhiyu  and
               Chen, Wenhu  and
               Smiley, Charese  and
               Shah, Sameena  and
               Borova, Iana  and
               Langdon, Dylan  and
               Moussa, Reema  and
               Beane, Matt  and
               Huang, Ting-Hao  and
               Routledge, Bryan  and
               Wang, William Yang},
  editor    = {Moens, Marie-Francine  and
               Huang, Xuanjing  and
               Specia, Lucia  and
               Yih, Scott Wen-tau},
  booktitle = {Proceedings of the 2021 Conference on Empirical Methods in Natural Language Processing},
  month     = nov,
  year      = {2021},
  address   = {Online and Punta Cana, Dominican Republic},
  publisher = {Association for Computational Linguistics},
  url       = {https://aclanthology.org/2021.emnlp-main.300/},
  doi       = {10.18653/v1/2021.emnlp-main.300},
  pages     = {3697--3711}
}

@article{zhu2021tat,
  title   = {TAT-QA: A question answering benchmark on a hybrid of tabular and textual content in finance},
  author  = {Zhu, Fengbin and Lei, Wenqiang and Huang, Youcheng and Wang, Chao and Zhang, Shuo and Lv, Jiancheng and Feng, Fuli and Chua, Tat-Seng},
  journal = {arXiv preprint arXiv:2105.07624},
  year    = {2021}
}

@article{ravichander2019question,
  title   = {Question answering for privacy policies: Combining computational and legal perspectives},
  author  = {Ravichander, Abhilasha and Black, Alan W and Wilson, Shomir and Norton, Thomas and Sadeh, Norman},
  journal = {arXiv preprint arXiv:1911.00841},
  year    = {2019}
}

@article{zhang2022opt,
  title         = {OPT: Open Pre-trained Transformer Language Models},
  author        = {Susan Zhang and Stephen Roller and Naman Goyal and Mikel Artetxe and Moya Chen and Shuohui Chen and Christopher Dewan and Mona Diab and Xian Li and Xi Victoria Lin and Todor Mihaylov and Myle Ott and Sam Shleifer and Kurt Shuster and Daniel Simig and Punit Singh Koura and Anjali Sridhar and Tianlu Wang and Luke Zettlemoyer},
  year          = {2022},
  eprint        = {2205.01068},
  archiveprefix = {arXiv},
  primaryclass  = {cs.CL},
  url           = {https://arxiv.org/abs/2205.01068}
}

@misc{allal2025smollm2smolgoesbig,
  title         = {SmolLM2: When Smol Goes Big -- Data-Centric Training of a Small Language Model},
  author        = {Loubna Ben Allal and Anton Lozhkov and Elie Bakouch and Gabriel Martín Blázquez and Guilherme Penedo and Lewis Tunstall and Andrés Marafioti and Hynek Kydlíček and Agustín Piqueres Lajarín and Vaibhav Srivastav and Joshua Lochner and Caleb Fahlgren and Xuan-Son Nguyen and Clémentine Fourrier and Ben Burtenshaw and Hugo Larcher and Haojun Zhao and Cyril Zakka and Mathieu Morlon and Colin Raffel and Leandro von Werra and Thomas Wolf},
  year          = {2025},
  eprint        = {2502.02737},
  archiveprefix = {arXiv},
  primaryclass  = {cs.CL},
  url           = {https://arxiv.org/abs/2502.02737}
}

@misc{jetsonorinnx,
  author       = {NVIDIA},
  howpublished = {\url{https://developer.download.nvidia.com/assets/embedded/secure/jetson/orin_nx/docs/Jetson_Orin_NX_DS-10712-001_v0.5.pdf}},
  title        = {NVIDIA Jetson Orin NX Series},
  year         = {2022}
}

@article{paszke2019pytorch,
  author    = {Paszke, Adam and Gross, Sam and Massa, Francisco and Lerer, Adam and Bradbury, James and Chanan, Gregory and Killeen, Trevor and Lin, Zeming and Gimelshein, Natalia and Antiga, Luca and Desmaison, Alban and Kopf, Andreas and Yang, Edward and DeVito, Zachary and Raison, Martin and Tejani, Alykhan and Chilamkurthy, Sasank and Steiner, Benoit and Fang, Lu and Bai, Junjie and Chintala, Soumith},
  booktitle = {Advances in Neural Information Processing Systems},
  editor    = {H. Wallach and H. Larochelle and A. Beygelzimer and F. d\textquotesingle Alch\'{e}-Buc and E. Fox and R. Garnett},
  pages     = {},
  publisher = {Curran Associates, Inc.},
  title     = {PyTorch: An Imperative Style, High-Performance Deep Learning Library},
  url       = {https://proceedings.neurips.cc/paper_files/paper/2019/file/bdbca288fee7f92f2bfa9f7012727740-Paper.pdf},
  volume    = {32},
  year      = {2019}
}

@misc{chetlur2014cudnn,
  title         = {cuDNN: Efficient Primitives for Deep Learning},
  author        = {Sharan Chetlur and Cliff Woolley and Philippe Vandermersch and Jonathan Cohen and John Tran and Bryan Catanzaro and Evan Shelhamer},
  year          = {2014},
  eprint        = {1410.0759},
  archiveprefix = {arXiv},
  primaryclass  = {cs.NE},
  url           = {https://arxiv.org/abs/1410.0759}
}

@article{douze2024faiss,
  title   = {The faiss library},
  author  = {Douze, Matthijs and Guzhva, Alexandr and Deng, Chengqi and Johnson, Jeff and Szilvasy, Gergely and Mazar{\'e}, Pierre-Emmanuel and Lomeli, Maria and Hosseini, Lucas and J{\'e}gou, Herv{\'e}},
  journal = {arXiv preprint arXiv:2401.08281},
  year    = {2024}
}

@article{malkov2020efficient,
  author   = {Malkov, Yu A. and Yashunin, D. A.},
  journal  = {IEEE Transactions on Pattern Analysis and Machine Intelligence},
  title    = {Efficient and Robust Approximate Nearest Neighbor Search Using Hierarchical Navigable Small World Graphs},
  year     = {2020},
  volume   = {42},
  number   = {4},
  pages    = {824-836},
  doi      = {10.1109/TPAMI.2018.2889473}
}

@article{lewis2020retrieval,
  author    = {Lewis, Patrick and Perez, Ethan and Piktus, Aleksandra and Petroni, Fabio and Karpukhin, Vladimir and Goyal, Naman and K\"{u}ttler, Heinrich and Lewis, Mike and Yih, Wen-tau and Rockt\"{a}schel, Tim and Riedel, Sebastian and Kiela, Douwe},
  booktitle = {Advances in Neural Information Processing Systems},
  editor    = {H. Larochelle and M. Ranzato and R. Hadsell and M.F. Balcan and H. Lin},
  pages     = {9459--9474},
  publisher = {Curran Associates, Inc.},
  title     = {Retrieval-Augmented Generation for Knowledge-Intensive NLP Tasks},
  url       = {https://proceedings.neurips.cc/paper_files/paper/2020/file/6b493230205f780e1bc26945df7481e5-Paper.pdf},
  volume    = {33},
  year      = {2020}
}

@article{vaswani2017attention,
  title   = {Attention is all you need},
  author  = {Vaswani, Ashish and Shazeer, Noam and Parmar, Niki and Uszkoreit, Jakob and Jones, Llion and Gomez, Aidan N and Kaiser, {\L}ukasz and Polosukhin, Illia},
  journal = {Advances in neural information processing systems},
  volume  = {30},
  year    = {2017}
}

@article{dao2022flashattention,
  title   = {Flashattention: Fast and memory-efficient exact attention with io-awareness},
  author  = {Dao, Tri and Fu, Dan and Ermon, Stefano and Rudra, Atri and R{\'e}, Christopher},
  journal = {Advances in neural information processing systems},
  volume  = {35},
  pages   = {16344--16359},
  year    = {2022}
}

@article{dao2023flashattention,
  title   = {Flashattention-2: Faster attention with better parallelism and work partitioning},
  author  = {Dao, Tri},
  journal = {arXiv preprint arXiv:2307.08691},
  year    = {2023}
}

@inproceedings{pan2025instattention,
  title        = {InstAttention: in-storage attention offloading for cost-effective long-context LLM inference},
  author       = {Pan, Xiurui and Li, Endian and Li, Qiao and Liang, Shengwen and Shan, Yizhou and Zhou, Ke and Luo, Yingwei and Wang, Xiaolin and Zhang, Jie},
  booktitle    = {2025 IEEE International Symposium on High Performance Computer Architecture (HPCA)},
  pages        = {1510--1525},
  year         = {2025},
  organization = {IEEE}
}

@inproceedings{jang2026hilos,
  author    = {Jang, Hongsun and Song, Jaeyong and Shin, Changmin and Noh, Si Ung and Jung, Jaewon and Park, Jisung and Lee, Jinho},
  title     = {A Cost-Effective Near-Storage Processing Solution for Offline Inference of Long-Context LLMs},
  year      = {2026},
  isbn      = {9798400723599},
  publisher = {Association for Computing Machinery},
  address   = {New York, NY, USA},
  url       = {https://doi.org/10.1145/3779212.3790119},
  doi       = {10.1145/3779212.3790119},
  pages     = {5--24},
  numpages  = {20},
  location  = {USA},
  series    = {ASPLOS '26}
}

@inproceedings{lee2025aif,
  title     = {AiF: Accelerating On-Device LLM Inference Using In-Flash Processing},
  author    = {Lee, Jaeyong and Kim, Hyeunjoo and Oh, Sanghun and Chun, Myoungjun and Kim, Myungsuk and Kim, Jihong},
  booktitle = {Proceedings of the 52nd Annual International Symposium on Computer Architecture},
  pages     = {529--543},
  year      = {2025}
}

@inproceedings{kim2025llmonthepalm,
  author    = {Kim, Hyunjin and Han, Sanghyeok and Kim, Jae-Joon},
  booktitle = {2025 IEEE/ACM International Conference On Computer Aided Design (ICCAD)},
  title     = {LLM-on-the-Palm: Mobile LLM Inference with PIM-Enhanced NAND Flash Memory},
  year      = {2025},
  volume    = {},
  number    = {},
  pages     = {1-9},
  doi       = {10.1109/ICCAD66269.2025.11240826}
}

@inproceedings{dac2016chisel,
  author    = {Bachrach, Jonathan and Vo, Huy and Richards, Brian and Lee, Yunsup and Waterman, Andrew and Avi\v{z}ienis, Rimas and Wawrzynek, John and Asanovi{\'c}, Krste},
  title     = {Chisel: Constructing Hardware in a Scala Embedded Language},
  booktitle = {Proceedings of the 53rd Annual Design Automation Conference (DAC)},
  year      = {2016},
  pages     = {1--10},
  doi       = {10.1145/2897937.2905004}
}

@article{lee2015hwacha,
  title   = {The Hwacha vector-fetch architecture manual, version 3.8. 1},
  author  = {Lee, Yunsup and Schmidt, Colin and Ou, Albert and Waterman, Andrew and Asanovic, K},
  journal = {EECS Department, University of California, Berkeley, Tech. Rep. UCB/EECS-2015-262},
  year    = {2015}
}

@inproceedings{gemmini-dac,
  author    = {Genc, Hasan and Kim, Seah and Amid, Alon and Haj-Ali, Ameer and Iyer, Vighnesh and Prakash, Pranav and Zhao, Jerry and Grubb, Daniel and Liew, Harrison and Mao, Howard and Ou, Albert and Schmidt, Colin and Steffl, Samuel and Wright, John and Stoica, Ion and Ragan-Kelley, Jonathan and Asanovic, Krste and Nikolic, Borivoje and Shao, Yakun Sophia},
  booktitle = {Proceedings of the 58th Annual Design Automation Conference (DAC)},
  title     = {Gemmini: Enabling Systematic Deep-Learning Architecture Evaluation via Full-Stack Integration},
  year      = {2021},
  volume    = {},
  number    = {},
  pages     = {}
}

@article{jung2017simplessd,
  title     = {SimpleSSD: Modeling solid state drives for holistic system simulation},
  author    = {Jung, Myoungsoo and Zhang, Jie and Abulila, Ahmed and Kwon, Miryeong and Shahidi, Narges and Shalf, John and Kim, Nam Sung and Kandemir, Mahmut},
  journal   = {IEEE Computer Architecture Letters},
  volume    = {17},
  number    = {1},
  pages     = {37--41},
  year      = {2017},
  publisher = {IEEE}
}

@inproceedings{yao2025cacheblend,
  title     = {Cacheblend: Fast large language model serving for rag with cached knowledge fusion},
  author    = {Yao, Jiayi and Li, Hanchen and Liu, Yuhan and Ray, Siddhant and Cheng, Yihua and Zhang, Qizheng and Du, Kuntai and Lu, Shan and Jiang, Junchen},
  booktitle = {Proceedings of the twentieth European conference on computer systems},
  pages     = {94--109},
  year      = {2025}
}

@misc{u200,
  author       = {AMD Xilinx},
  howpublished = {\url{https://www.amd.com/content/dam/amd/en/documents/products/accelerators/alveo/alveo-product-brief.pdf}},
  title        = {AMD ALVEO U200 U250: Adaptable Accelerator Cards for Data Center Workloads},
  year         = {2023}
}

@misc{cuda,
  title        = {CUDA Toolkit Documentation},
  author       = {{NVIDIA Corporation}},
  year         = {2025},
  howpublished = {\url{https://developer.nvidia.com/cuda-toolkit}},
  note         = {Accessed: 2025-08-20}
}

@article{alkhalaf2024applying,
  title     = {Applying generative AI with retrieval augmented generation to summarize and extract key clinical information from electronic health records},
  author    = {Alkhalaf, Mohammad and Yu, Ping and Yin, Mengyang and Deng, Chao},
  journal   = {Journal of biomedical informatics},
  volume    = {156},
  pages     = {104662},
  year      = {2024},
  publisher = {Elsevier}
}

@misc{wu2024medicalgraphragsafe,
  title         = {Medical Graph RAG: Towards Safe Medical Large Language Model via Graph Retrieval-Augmented Generation},
  author        = {Junde Wu and Jiayuan Zhu and Yunli Qi and Jingkun Chen and Min Xu and Filippo Menolascina and Vicente Grau},
  year          = {2024},
  eprint        = {2408.04187},
  archiveprefix = {arXiv},
  primaryclass  = {cs.CV},
  url           = {https://arxiv.org/abs/2408.04187}
}

@article{amugongo2025retrieval,
  title     = {Retrieval augmented generation for large language models in healthcare: A systematic review},
  author    = {Amugongo, Lameck Mbangula and Mascheroni, Pietro and Brooks, Steven and Doering, Stefan and Seidel, Jan},
  journal   = {PLOS Digital Health},
  volume    = {4},
  number    = {6},
  pages     = {e0000877},
  year      = {2025},
  publisher = {Public Library of Science San Francisco, CA USA}
}

@inproceedings{choi2025finder,
  title     = {Finder: Financial dataset for question answering and evaluating retrieval-augmented generation},
  author    = {Choi, Chanyeol and Kwon, Jihoon and Ha, Jaeseon and Choi, Hojun and Kim, Chaewoon and Lee, Yongjae and Sohn, Jy-yong and Lopez-Lira, Alejandro},
  booktitle = {Proceedings of the 6th ACM International Conference on AI in Finance},
  pages     = {638--646},
  year      = {2025}
}

@article{kim2025optimizing,
  title   = {Optimizing retrieval strategies for financial question answering documents in retrieval-augmented generation systems},
  author  = {Kim, Sejong and Song, Hyunseo and Seo, Hyunwoo and Kim, Hyunjun},
  journal = {arXiv preprint arXiv:2503.15191},
  year    = {2025}
}

@article{iaroshev2024evaluating,
  title     = {Evaluating retrieval-augmented generation models for financial report question and answering},
  author    = {Iaroshev, Ivan and Pillai, Ramalingam and Vaglietti, Leandro and Hanne, Thomas},
  journal   = {Applied Sciences},
  volume    = {14},
  number    = {20},
  pages     = {9318},
  year      = {2024},
  publisher = {MDPI}
}

@inproceedings{wang2024mememo,
  title     = {Mememo: On-device retrieval augmentation for private and personalized text generation},
  author    = {Wang, Zijie J and Chau, Duen Horng},
  booktitle = {Proceedings of the 47th International ACM SIGIR Conference on Research and Development in Information Retrieval},
  pages     = {2765--2770},
  year      = {2024}
}

@inproceedings{qin2024robust,
  author    = {Qin, Ruiyang and Yan, Zheyu and Zeng, Dewen and Jia, Zhenge and Liu, Dancheng and Liu, Jianbo and Abbasi, Ahmed and Zheng, Zhi and Cao, Ningyuan and Ni, Kai and Xiong, Jinjun and Shi, Yiyu},
  title     = {Robust Implementation of Retrieval-Augmented Generation on Edge-based Computing-in-Memory Architectures},
  year      = {2025},
  isbn      = {9798400710773},
  publisher = {Association for Computing Machinery},
  address   = {New York, NY, USA},
  url       = {https://doi.org/10.1145/3676536.3676674},
  doi       = {10.1145/3676536.3676674},
  booktitle = {Proceedings of the 43rd IEEE/ACM International Conference on Computer-Aided Design},
  articleno = {50},
  numpages  = {9},
  location  = {Newark Liberty International Airport Marriott, New York, NY, USA},
  series    = {ICCAD '24}
}

@inproceedings{mahapatra2025storage,
  title     = {In-storage acceleration of retrieval augmented generation as a service},
  author    = {Mahapatra, Rohan and Santhanam, Harsha and Priebe, Christopher and Xu, Hanyang and Esmaeilzadeh, Hadi},
  booktitle = {Proceedings of the 52nd Annual International Symposium on Computer Architecture},
  pages     = {450--466},
  year      = {2025}
}

@article{loo2025mobilerag,
  title   = {Mobilerag: Enhancing mobile agent with retrieval-augmented generation},
  author  = {Loo, Gowen and Liu, Chang and Yin, Qinghong and Chen, Xiang and Chen, Jiawei and Zhang, Jingyuan and Tian, Yu},
  journal = {arXiv preprint arXiv:2509.03891},
  year    = {2025}
}

@misc{liu2025lmcacheefficientkvcache,
  title         = {LMCache: An Efficient KV Cache Layer for Enterprise-Scale LLM Inference},
  author        = {Yuhan Liu and Yihua Cheng and Jiayi Yao and Yuwei An and Xiaokun Chen and Shaoting Feng and Yuyang Huang and Samuel Shen and Rui Zhang and Kuntai Du and Junchen Jiang},
  year          = {2025},
  eprint        = {2510.09665},
  archiveprefix = {arXiv},
  primaryclass  = {cs.LG},
  url           = {https://arxiv.org/abs/2510.09665}
}

@article{gim2024prompt,
  title   = {Prompt cache: Modular attention reuse for low-latency inference},
  author  = {Gim, In and Chen, Guojun and Lee, Seung-seob and Sarda, Nikhil and Khandelwal, Anurag and Zhong, Lin},
  journal = {Proceedings of Machine Learning and Systems},
  volume  = {6},
  pages   = {325--338},
  year    = {2024}
}

@article{pan2024instinfer,
  title   = {Instinfer: In-storage attention offloading for cost-effective long-context llm inference},
  author  = {Pan, Xiurui and Li, Endian and Li, Qiao and Liang, Shengwen and Shan, Yizhou and Zhou, Ke and Luo, Yingwei and Wang, Xiaolin and Zhang, Jie},
  journal = {arXiv preprint arXiv:2409.04992},
  year    = {2024}
}

@article{10.1145/3773772,
  author    = {Qin, Ruoyu and Li, Zheming and He, Weiran and Cui, Jialei and Tang, Heyi and Ren, Feng and Ma, Teng and Cai, Shangming and Zhang, Yineng and Zhang, Mingxing and Wu, Yongwei and Zheng, Weimin and Xu, Xinran},
  title     = {Mooncake: A KVCache-centric Disaggregated Architecture for LLM Serving},
  year      = {2025},
  publisher = {Association for Computing Machinery},
  address   = {New York, NY, USA},
  issn      = {1553-3077},
  url       = {https://doi.org/10.1145/3773772},
  doi       = {10.1145/3773772},
  note      = {Just Accepted},
  journal   = {ACM Trans. Storage},
  month     = nov
}

@inproceedings{ren2025characterizing,
  title     = {An i/o characterizing study of offloading llm models and kv caches to nvme ssd},
  author    = {Ren, Zebin and Doekemeijer, Krijn and De Matteis, Tiziano and Pinto, Christian and Stoica, Radu and Trivedi, Animesh},
  booktitle = {Proceedings of the 5th Workshop on Challenges and Opportunities of Efficient and Performant Storage Systems},
  pages     = {23--33},
  year      = {2025}
}

@misc{holmes2024deepspeedfastgenhighthroughputtextgeneration,
  title         = {DeepSpeed-FastGen: High-throughput Text Generation for LLMs via MII and DeepSpeed-Inference},
  author        = {Connor Holmes and Masahiro Tanaka and Michael Wyatt and Ammar Ahmad Awan and Jeff Rasley and Samyam Rajbhandari and Reza Yazdani Aminabadi and Heyang Qin and Arash Bakhtiari and Lev Kurilenko and Yuxiong He},
  year          = {2024},
  eprint        = {2401.08671},
  archiveprefix = {arXiv},
  primaryclass  = {cs.PF},
  url           = {https://arxiv.org/abs/2401.08671}
}

@inproceedings{pmlr-v202-sheng23a,
  title     = {{F}lex{G}en: High-Throughput Generative Inference of Large Language Models with a Single {GPU}},
  author    = {Sheng, Ying and Zheng, Lianmin and Yuan, Binhang and Li, Zhuohan and Ryabinin, Max and Chen, Beidi and Liang, Percy and Re, Christopher and Stoica, Ion and Zhang, Ce},
  booktitle = {Proceedings of the 40th International Conference on Machine Learning},
  pages     = {31094--31116},
  year      = {2023},
  editor    = {Krause, Andreas and Brunskill, Emma and Cho, Kyunghyun and Engelhardt, Barbara and Sabato, Sivan and Scarlett, Jonathan},
  volume    = {202},
  series    = {Proceedings of Machine Learning Research},
  month     = {23--29 Jul},
  publisher = {PMLR},
  url       = {https://proceedings.mlr.press/v202/sheng23a.html}
}

@article{cordonnier2020multi,
  title   = {Multi-head attention: Collaborate instead of concatenate},
  author  = {Cordonnier, Jean-Baptiste and Loukas, Andreas and Jaggi, Martin},
  journal = {arXiv preprint arXiv:2006.16362},
  year    = {2020}
}

@inproceedings{ainslie-etal-2023-gqa,
  title     = {{GQA}: Training Generalized Multi-Query Transformer Models from Multi-Head Checkpoints},
  author    = {Ainslie, Joshua  and
               Lee-Thorp, James  and
               de Jong, Michiel  and
               Zemlyanskiy, Yury  and
               Lebron, Federico  and
               Sanghai, Sumit},
  editor    = {Bouamor, Houda  and
               Pino, Juan  and
               Bali, Kalika},
  booktitle = {Proceedings of the 2023 Conference on Empirical Methods in Natural Language Processing},
  month     = dec,
  year      = {2023},
  address   = {Singapore},
  publisher = {Association for Computational Linguistics},
  url       = {https://aclanthology.org/2023.emnlp-main.298/},
  doi       = {10.18653/v1/2023.emnlp-main.298},
  pages     = {4895--4901}
}

@article{schroeder2025vcache,
  title   = {vcache: Verified semantic prompt caching},
  author  = {Schroeder, Luis Gaspar and Desai, Aditya and Cuadron, Alejandro and Chu, Kyle and Liu, Shu and Zhao, Mark and Krusche, Stephan and Kemper, Alfons and Zaharia, Matei and Gonzalez, Joseph E},
  journal = {arXiv preprint arXiv:2502.03771},
  year    = {2025}
}

@misc{zhu2024efficientpromptcachingembedding,
  title         = {Efficient Prompt Caching via Embedding Similarity},
  author        = {Hanlin Zhu and Banghua Zhu and Jiantao Jiao},
  year          = {2024},
  eprint        = {2402.01173},
  archiveprefix = {arXiv},
  primaryclass  = {cs.CL},
  url           = {https://arxiv.org/abs/2402.01173}
}

@inproceedings{sharma2013greencache,
  title     = {GreenCache: augmenting off-the-grid cellular towers with multimedia caches},
  author    = {Sharma, Navin and Krishnappa, Dilip Kumar and Irwin, David and Zink, Michael and Shenoy, Prashant},
  booktitle = {Proceedings of the 4th ACM multimedia systems conference},
  pages     = {271--280},
  year      = {2013}
}

@article{yang2025cacheclip,
  title   = {CacheClip: Accelerating RAG with Effective KV Cache Reuse},
  author  = {Yang, Bin and Leng, Qiuyu and Zeng, Jun and Wu, Zhenhua},
  journal = {arXiv preprint arXiv:2510.10129},
  year    = {2025}
}

@article{edge2024local,
  title   = {From local to global: A graph rag approach to query-focused summarization},
  author  = {Edge, Darren and Trinh, Ha and Cheng, Newman and Bradley, Joshua and Chao, Alex and Mody, Apurva and Truitt, Steven and Metropolitansky, Dasha and Ness, Robert Osazuwa and Larson, Jonathan},
  journal = {arXiv preprint arXiv:2404.16130},
  year    = {2024}
}

@article{zhuang2025linearrag,
  title   = {Linearrag: Linear graph retrieval augmented generation on large-scale corpora},
  author  = {Zhuang, Luyao and Chen, Shengyuan and Xiao, Yilin and Zhou, Huachi and Zhang, Yujing and Chen, Hao and Zhang, Qinggang and Huang, Xiao},
  journal = {arXiv preprint arXiv:2510.10114},
  year    = {2025}
}

@inproceedings{nogueira2020document,
  title     = {Document ranking with a pretrained sequence-to-sequence model},
  author    = {Nogueira, Rodrigo and Jiang, Zhiying and Pradeep, Ronak and Lin, Jimmy},
  booktitle = {Findings of the association for computational linguistics: EMNLP 2020},
  pages     = {708--718},
  year      = {2020}
}

@inproceedings{trivedi2023interleaving,
  title     = {Interleaving retrieval with chain-of-thought reasoning for knowledge-intensive multi-step questions},
  author    = {Trivedi, Harsh and Balasubramanian, Niranjan and Khot, Tushar and Sabharwal, Ashish},
  booktitle = {Proceedings of the 61st annual meeting of the association for computational linguistics (volume 1: long papers)},
  pages     = {10014--10037},
  year      = {2023}
}

@article{li2024snapkv,
  title   = {Snapkv: Llm knows what you are looking for before generation},
  author  = {Li, Yuhong and Huang, Yingbing and Yang, Bowen and Venkitesh, Bharat and Locatelli, Acyr and Ye, Hanchen and Cai, Tianle and Lewis, Patrick and Chen, Deming},
  journal = {Advances in Neural Information Processing Systems},
  volume  = {37},
  pages   = {22947--22970},
  year    = {2024}
}

@article{kim2026kvzip,
  title   = {Kvzip: Query-agnostic kv cache compression with context reconstruction},
  author  = {Kim, Jang-Hyun and Kim, Jinuk and Kwon, Sangwoo and Lee, Jae W and Yun, Sangdoo and Song, Hyun Oh},
  journal = {Advances in Neural Information Processing Systems},
  volume  = {38},
  pages   = {167563--167591},
  year    = {2026}
}

@article{muralimanohar2009cacti,
  title   = {CACTI 6.0: A tool to model large caches},
  author  = {Muralimanohar, Naveen and Balasubramonian, Rajeev and Jouppi, Norman P},
  journal = {HP laboratories},
  volume  = {27},
  pages   = {28},
  year    = {2009}
}

@inproceedings{alizadeh2024llm,
  title     = {Llm in a flash: Efficient large language model inference with limited memory},
  author    = {Alizadeh, Keivan and Mirzadeh, Seyed Iman and Belenko, Dmitry and Khatamifard, S and Cho, Minsik and Del Mundo, Carlo C and Rastegari, Mohammad and Farajtabar, Mehrdad},
  booktitle = {Proceedings of the 62nd Annual Meeting of the Association for Computational Linguistics (Volume 1: Long Papers)},
  pages     = {12562--12584},
  year      = {2024}
}

@article{kuon2007measuring,
  title     = {Measuring the gap between {FPGAs} and {ASICs}},
  author    = {Kuon, Ian and Rose, Jonathan},
  journal   = {IEEE Transactions on Computer-Aided Design of Integrated Circuits and Systems},
  volume    = {26},
  number    = {2},
  pages     = {203--215},
  year      = {2007},
  publisher = {IEEE}
}

\end{document}